\documentclass[useAMS,usenatbib]{mn2e}
\usepackage{graphicx}
\usepackage{natbib}
\newcommand\lsim{~\lower.5ex\hbox{$\buildrel < \over \sim$}~}
\newcommand\gsim{~\lower.5ex\hbox{$\buildrel > \over \sim$}~}

\title[Axisymmetric modes in vertically stratified self-gravitating discs]
{Axisymmetric modes in vertically stratified self-gravitating discs}
\author[G.~R.~Mamatsashvili and W.~K.~M.~Rice] {G.~R.~Mamatsashvili$^{1,2}$\thanks{E-mail:
grm@roe.ac.uk} and W.~K.~M.~Rice
$^{1}$\\
$^{1}$ SUPA, Institute for Astronomy, University of Edinburgh,
Blackford Hill, Edinburgh EH9 3HJ, Scotland \\
$^{2}$ Georgian National Astrophysical Observatory, Ilia State
University, 2a Kazbegi Ave., Tbilisi 0160, Georgia}
\begin{document}

\date{Accepted 2010 April 09. Received 2010 April 08; in
original form 2010 February 09}

\pagerange{\pageref{firstpage}--\pageref{lastpage}} \pubyear{2010}

\maketitle

\label{firstpage}

\begin{abstract}
We carry out a linear analysis of the vertical normal modes of
axisymmetric perturbations in stratified, compressible,
self-gravitating gaseous discs in the shearing box approximation. An
unperturbed disc has a polytropic vertical structure that allows us
to study specific dynamics for subadiabatic, adiabatic and
superadiabatic vertical stratifications, by simply varying the
polytropic index. In the absence of self-gravity, four well-known
principal modes can be identified in a stratified disc: acoustic
p-modes, surface gravity f-modes, buoyancy g-modes and inertial
r-modes. After classifying and characterizing modes in the
non-self-gravitating case, we include self-gravity in the
perturbation equations and in the equilibrium and investigate how it
modifies the properties of these four modes. We find that
self-gravity, to a certain degree, reduces their frequencies and
changes the structure of the dispersion curves and eigenfunctions at
radial wavelengths comparable to the disc height. Its influence on
the basic branch of the r-mode, in the case of subadiabatic and
adiabatic stratifications, and on the basic branch of the g-mode, in
the case of superadiabatic stratification (which in addition
exhibits convective instability), does appear to be strongest.
Reducing the three-dimensional Toomre's parameter $Q_{3D}$ results
in the latter modes becoming unstable due to self-gravity, so that
they determine the onset criterion and nature of the gravitational
instability of a vertically stratified disc. By contrast, the p-, f-
and convectively stable g-modes, although their corresponding
$\omega^2$ are reduced by self-gravity, never become unstable
however small the value of $Q_{3D}$. This is a consequence of the
three-dimensionality of the disc. The eigenfunctions corresponding
to the gravitationally unstable modes are intrinsically
three-dimensional. We also contrast the more exact instability
criterion based on our three-dimensional model with that of density
waves in two-dimensional (razor-thin) discs. Based on these
findings, we comment on the origin of surface distortions seen in
numerical simulations of self-gravitating discs.
\end{abstract}


\section{Introduction}

Self-gravity plays an important role in a variety of astrophysical
systems. It is a main agent determining the dynamical evolution of
star clusters, galaxies, various types of accretion discs, etc.
Particularly in protoplanetary discs, that are the central subject
of our study, self-gravity provides one of the main source of
outward angular momentum transport through the excitation of density
waves \citep{PS91,LB94,LR04,LR05} and is able to cause fragmentation
of a disc into bound clumps, or planets, under certain conditions
\citep{Gam01,Riceetal03,RLA05,B04,Mayeretal07,Raf07}. Starting with
the seminal paper by \citet{T64}, there have been a large number of
studies of the stability of self-gravitating gaseous discs, both in
the linear
\citep[e.g.,][]{GLB65a,GLB65b,GT78,ARS89,Bertetal89,PL89,LKA97} and
non-linear regimes, including other relevant physical factors (e.g.,
heating, cooling, radiation transport) with up-to-date numerical
techniques
\citep[e.g.,][]{PS91,LB94,B98,Petal00,Petal03,Gam01,B03,JG03,Riceetal03,RLA05,B04,Metal05,Betal06,Mayeretal07,
SW09}.

Linear stability analysis in a vast majority of cases is restricted,
for simplicity, to razor-thin, or two-dimensional (2D) discs that
are obtained by vertically averaging all quantities. In other words,
perturbations are assumed to have large horizontal scales compared
with the disc thickness. In this limit, the well-known Toomre's
parameter
$$
Q_{2D}=\frac{c_s\Omega}{\pi G \Sigma}
$$
controls the stability of self-gravitating discs \citep{T64}. In
this case, a density wave, which is the only mode in a 2D disc, is
influenced by self-gravity and thus can become unstable, as the
local dispersion relation for the latter clearly demonstrates
\citep{GT78,BT87,Bertetal89}.

Stability analysis in a more realistic case of self-gravitating
three-dimensional (3D) discs is more complicated. The disc is
vertically stratified due to both its own self-gravity and the
vertical component of the gravity of a central object. Depending on
the nature of the stratification, there exists a whole new set of
various vertical modes in the system (see below), some of which can
become unstable due to self-gravity on horizontal length scales
comparable to the disc thickness. In this situation, the vertical
variation of perturbations is important and for a correct
characterization of the gravitational instability it is necessary to
introduce another parameter not involving height-dependent
variables, such as the sound speed in Toomre's parameter.
Furthermore, not all types of stratification permit two-dimensional
modes, that is, modes with no vertical motions commonly occurring in
the 2D treatment. For example, in non-self-gravitating discs with
polytropic vertical structure, there are no 2D modes
\citep{LPS90,LP93a,KP95} implying that the dynamics does not always
reduce to that of the 2D case. Therefore, a more accurate stability
analysis of self-gravitating discs should necessarily be
three-dimensional.

Obviously, before studying the gravitational instability of
stratified discs, one must first classify and characterize vertical
normal modes of perturbations in the simplified case of no
self-gravity. Analysis of the modal structure of stratified,
polytropic, compressible, non-self-gravitating discs has been done
in several papers: \citet[][hereafter RPL]{RPL88},
\citet[][hereafter KP]{KP95}, \citet{Og98}. In convectively stable
discs, i.e., with subadiabatic vertical stratification, four
principal types of vertical modes can be distinguished. These modes
are: acoustic p-modes, surface gravity f-modes, buoyancy g-modes and
inertial r-modes. The modes are named after their corresponding
restoring forces, which can be well identified for each mode at
horizontal wavelengths smaller than the disc height and are provided
by one of the following: compressibility/pressure, displacements of
free surfaces of a disc, buoyancy due to vertical stratification and
inertial forces due to disc rotation, respectively, for the p-, f-,
g- and r-modes. In the case of superadiabatic stratification, the r-
and g-modes merge and appear as a single mode, which becomes
convectively unstable for horizontal wavenumbers larger than a
certain value (RPL); the p- and f-modes remain qualitatively
unchanged. For neutral/adiabatic stratification, buoyancy is absent
and the g-mode disappears. \emph{The main purpose of this paper is
to investigate how self-gravity modifies the frequencies and the
structure of the eigenfunctions of these modes, which mode acquires
the largest positive growth rate due to self-gravity and, therefore,
determines the onset criterion and nature of the gravitational
instability of a stratified disc.} So, the mode dynamics in the 3D
case can appear more complex than that in the 2D one, where only the
density wave mode can be subject to gravitational instability.
Previously, \citet[][hereafter GLB]{GLB65a} considered gravitational
instability in a uniformly rotating gaseous slab with an adiabatic
vertical stratification, thereby leaving out all modes associated
with buoyancy. Other studies also considered the gravitational
instability of 3D galactic discs, however, the analysis was
essentially 2D, finite-thickness effects were only taken into
account by means of various reduction factors in 2D dispersion
relations \citep{Shu68,Rom92,Rom94}. In all these studies, as in
GLB, the main focus was on finding the criterion for the onset of
gravitational instability, so that a full analysis of various types
of vertical normal modes existing in stratified self-gravitating
discs was not carried out. Actually, we generalize the study of GLB
to subadiabatic and superadiabatic stratifications having different
modal structure.

Another motivation for our study is that the f-mode is thought to
play an important dynamical role in self-gravitating discs. The
non-linear behaviour of 3D perturbations involving large surface
distortions, as seen in numerical simulations, has been attributed
to the surface gravity f-mode \citep{Petal00}. However, this was
done without analysing the behaviour of other vertical modes under
self-gravity. It was shown that the f-mode leads to a large energy
dissipation in the vicinity of the disc surface, which may
facilitate disc cooling, because the energy is deposited at smaller
optical depth where it can be radiated away more quickly \citep[see
e.g.,][]{JG03,Betal06}. Later it was realized that in fact the
non-linear vertical motions in self-gravitating discs can be much
more complex than just the f-mode and can have a shock character
\citep[shock bores,][]{BD06}. Thus, in the 3D case, the dynamics of
self-gravitating discs is much richer and diverse than that of
idealized 2D ones and requires further study. To fully understand
the origin of such three-dimensional effects and what type of
instability they are associated with, one should start with a
rigorous linear study of the characteristic properties of all the
types of vertical normal modes mentioned above, not only the f-mode,
in the presence of self-gravity. The present work is just a first
step in this direction.

Numerical simulations of self-gravitating discs are often in the
context of global discs
\citep[e.g.,][]{Petal00,Petal03,Riceetal03,RLA05,LR04,LR05,Betal06,Bol09}
and, therefore, are not always able to well resolve vertical
motions, which, as shown in the present study, inevitably arise
during the development of the gravitational instability associated
with intrinsically three-dimensional modes. So, these simulations
may not quite accurately capture all the subtleties of the
gravitational instability in 3D discs. In this connection, we should
mention the work by \citet{N06} that extensively discusses the issue
of vertical resolution and its importance in the outcome of the
gravitational instability in numerical simulations of
self-gravitation discs. Resolving and analysing vertical motions are
also crucial for properly understanding cooling processes in discs
and, particularly, whether convection is able to provide
sufficiently effective cooling for disc fragmentation to occur,
which is still a matter for debate in the literature
\citep{B04,Mayeretal07,Betal06,Betal07,Raf07}. In addition, these
studies, for simplicity, use the criterion for gravitational
instability based on the two-dimensional Toomre's parameter, which,
as we will demonstrate, cannot always be uniquely mapped into an
analogous three-dimensional stability parameter and give a precise
criterion for the onset of gravitational instability.

In this paper, following other works in a similar vein:
\citet[][hereafter LP]{LP93a}, KP, \citet[][hereafter LO98]{LO98},
we adopt the shearing box approximation and consider the linear
dynamics of vertical normal modes of perturbations in a
compressible, stratified, self-gravitating gaseous disc with
Keplerian rotation. In the unperturbed disc, pressure and density
are related by a polytropic law, which is a reasonably good
approximation for optically thick discs (see e.g., LO98). This
allows us to consider the specific features of the dynamics for
subadiabatic, adiabatic and superadiabatic vertical stratifications by simply
varying the polytropic index. As a first step towards understanding
the effects of self-gravity on the perturbation modes, we restrict
ourselves to axisymmetric perturbations only. The linear results
obtained here will form the basis for studying the non-linear
development of gravitational instability in the local shearing box
approximation that allows much higher numerical resolution than
global disc models can afford.

The paper is organized as follows. The physical model and basic
equations are introduced in Section 2. The classification of
vertical modes in the absence of self-gravity is performed in
Section 3. Effects of self-gravity on all normal modes are analysed
in Section 4. In Section 5, we focus on the properties of
gravitational instability in 3D. Comparison between the criteria of
gravitational instability in 3D and 2D is made in Section 6. Summary
and discussions are given in Section 7.

\section{Physical Model and Equations}

In order to study the dynamics of three-dimensional modes in gaseous
self-gravitating discs, following LP, KP, LO98, we adopt a local
shearing box approximation \citep{GLB65b}. In the shearing box
model, the disc dynamics is studied in a local Cartesian reference
frame rotating with the angular velocity of disc rotation at some
fiducial radius from the central star, so that curvature effects due
to cylindrical geometry of the disc are ignored. In this coordinate
frame, the unperturbed differential rotation of the disc manifests
itself as a parallel azimuthal flow with a constant velocity shear
in the radial direction. A Coriolis force is included to take into
account the effects of coordinate frame rotation. The vertical
component of the gravity force of the central object is also
present. As a result, we can write down the three-dimensional
shearing box equations
\begin{equation}
\frac{d{\bf u}}{dt}+2\Omega{\bf\hat{z}}\times {\bf
u}+\nabla\left(-q\Omega^2x^2+\frac{1}{2}\Omega^2z^2+\psi\right)+\frac{1}{\rho}\nabla
p=0,
\end{equation}
\begin{equation}
\frac{d\rho}{dt}+\rho{\bf \nabla}\cdot{\bf u}=0
\end{equation}
and the equation of conservation of entropy
\begin{equation}
\frac{d}{dt}\left(\frac{p}{\rho^{\gamma}}\right)=0.
\end{equation}
This set of equations is supplemented with Poisson's equation to
take care of disc self-gravity
\begin{equation}
\left(\frac{\partial^2}{\partial x^2}+\frac{\partial^2}{\partial
y^2}+\frac{\partial^2}{\partial z^2}\right)\psi=4\pi G\rho.
\end{equation}
Here ${\bf u}=(u_x,u_y,u_z)$ is the velocity in the local frame, $p,
\rho, \psi$ are, respectively, the pressure, density and the
gravitational potential of the disc gas. $\Omega$ is the angular
velocity of the local reference frame rotation as a whole, $x,y,z$
are, respectively, the radial, azimuthal and vertical coordinates,
${\bf \hat{z}}$ is the unit vector along the vertical direction and
$d/dt=\partial/\partial t+{\bf u}\cdot \nabla$ is the total time
derivative. The shear parameter is $q=1.5$ for the Keplerian
rotation considered in this paper. The adiabatic index, or the ratio
of specific heats, $\gamma=1.4$ as typical of a disc composed of
molecular hydrogen; we adopt this value throughout the paper.

\subsection{The equilibrium disc model}

Equations (1-4) have an equilibrium solution that is stationary and
axisymmetric. In this unperturbed state, the velocity field
represents, as noted above, a parallel azimuthal flow, ${\bf
u_{0}}$, with a constant radial shear $q$:
$$
u_{x0}=u_{z0}=0,~~u_{y0}=-q\Omega x.
$$
In the shearing box, the equilibrium density $\rho_0$, pressure
$p_0$ and gravitational potential $\psi_0$ depend only on the
vertical $z-$coordinate and satisfy the hydrostatic relation
\begin{equation}
g_0\equiv -\frac{1}{\rho_0}\frac{dp_{0}}{dz}=\Omega^2
z+\frac{d\psi_0}{dz},
\end{equation}
\begin{equation}
\frac{d^2\psi_0}{dz^2}=4\pi G \rho_0.
\end{equation}
As in LP, KP and RPL, pressure and density in the unperturbed disc
are related by a polytropic relationship of the form
\begin{equation}
p_0=K\rho_0^{1+1/s},
\end{equation}
where $K$ is the polytropic constant and $s>0$ is the polytropic
index. The Brunt-V\"{a}is\"{a}l\"{a} frequency squared is defined as
\begin{equation}
N_0^2\equiv
\frac{g_0}{\rho_0}\left(\frac{1}{c_{s}^2}\frac{dp_0}{dz}-\frac{d\rho_0}{dz}
\right)=\left(\frac{\gamma s}{s+1}-1\right)\frac{g_0^2}{c_{s}^2},
\end{equation} where $c_{s}^2=\gamma p_0/\rho_0$ is the adiabatic sound speed
squared. If $1+1/s<\gamma$ (subadiabatic thermal stratification),
then $N_0^2>0$ all along the height and the equilibrium vertical
structure of the disc is convectively stable. In the opposite case
$1+1/s>\gamma$ (superadiabatic thermal stratification), $N_0^2<0$
everywhere and this corresponds to a convectively unstable
equilibrium. For $1+1/s=\gamma$ (adiabatic thermal stratification),
$N_0^2=0$ and all the motions/modes due to buoyancy disappear. Later
we will consider these three regimes separately.

To determine the vertical structure, we need to solve equations
(5-6) subject to the boundary condition that the pressure vanish at
the free surface of the disc. Because we have a polytropic model, it
is convenient to work with the pseudo-enthalpy
\[
w_0=(s+1)K\rho_0^{1/s}.
\]
The disc structure is also symmetric with respect to the midplane,
$z=0$, and, as a consequence, it follows from equations (5-7) that
the derivative of $w_0$ at the midplane vanishes (i.e.,
$dw_0/dz(0)=0$). Because of this reflection symmetry, we consider
only the upper half of the disc, $z\geq 0$. At the surface of the
disc $w_0=0$, similar to the pressure. This allows us to determine
the equilibrium vertical structure of the disc. The non-dimensional
variables being used throughout the paper are:
\[
x\rightarrow \frac{x\Omega}{c_{sm}},~~~~y\rightarrow
\frac{y\Omega}{c_{sm}},~~~~z\rightarrow \frac{z\Omega}{c_{sm}},
\]
\[
\rho_0 \rightarrow \frac{\rho_0}{\rho_m},~~~w_0 \rightarrow
\frac{w_0}{w_m},~~~c_s \rightarrow \frac{c_s}{c_{sm}},
\]
where $\rho_m\equiv \rho_0(0),~w_m\equiv w_0(0),~c_{sm}\equiv
c_s(0)$ are the midplane values of the equilibrium density,
pseudo-enthalpy and sound speed. We define the three-dimensional
analogue of Toomre's parameter as
$$
Q_{3D}=\frac{\Omega^2}{4\pi G\rho_m},
$$
which is a measure of disc self-gravity (from now on until Section
6, we will use $Q_{3D}$ without subscript everywhere, so it should
not be confused with the 2D Toomre's parameter). Using that at the
midplane $dw_0/dz(0)=0$, we can derive from equations (5-7) a single
equation for the normalized pseudo-enthalpy
\begin{equation}
\frac{s+1}{2\gamma}\left(\frac{dw_0}{dz}\right)^2=(1-w_0)+\frac{1}{Q(s+1)}(1-w_0^{s+1}).
\end{equation}
The normalized density and sound speed are found from
$\rho_0=w_0^s,~c_s^2=w_0$. Equation (9) shows that even though the
disc is in Keplerian rotation (i.e., self-gravity in the radial
direction can be neglected), it must be included in determining the
vertical structure. We integrate equation (9), starting at $z=0$
with $w_0(0)=1$, until $w_0$, monotonically decreasing with $z$,
reaches zero at some finite height/edge $z=h$, which is therefore
determined as a result of the integration process. At this edge, the
density and sound speed also vanish, $\rho_0(h)=0, c_s(h)=0$ (see
also GLB and RPL). The height and entire vertical structure of a
polytropic disc are therefore uniquely specified solely by $Q$ and
$s$, which are free parameters in equation (9).
\begin{figure}
\centering\includegraphics[width=\columnwidth]{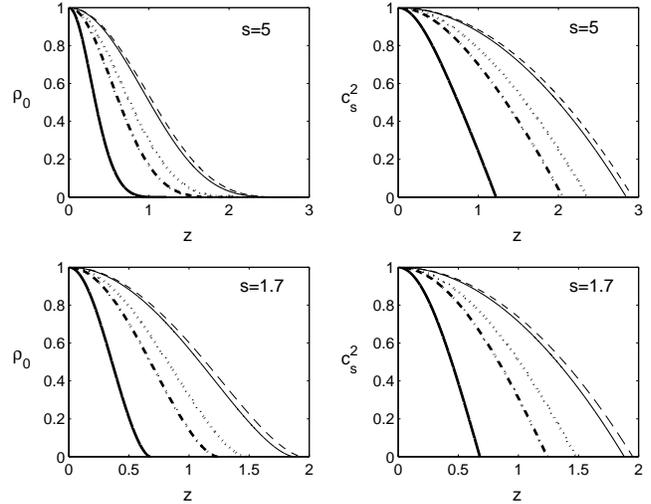}
\caption{Vertical variations of the normalized density and squared
sound speed for subadiabatic ($s=5$) and superadiabatic ($s=1.7$)
equilibrium states with $Q=0.1$ (thick solid lines), $Q=0.5$
(dashed-dotted lines), $Q=1$ (dotted lines), $Q=10$ (thin solid
lines) and the non-self-gravitating case ($Q\rightarrow \infty$,
dashed lines). The lines corresponding to $Q=10$ are close to those
in the non-self-gravitating limit, implying that the role of
self-gravity is negligible for $Q\geq 10$. The disc heights, $h$,
are different depending on $Q$ and $s$.}
\end{figure}

Figure 1 illustrates the sub- and superadiabatic equilibrium
vertical structures obtained from equation (9) for various $Q$. For
a fixed $s$, the disc height $h$ decreases with decreasing $Q$. The
sound speed is a decreasing function of $z$ that does not permit the
existence of two-dimensional modes in three-dimensional models of
polytropic discs as opposed to isothermal ones \citep[][LP,
KP]{LPS90}. We also see that the equilibrium structures for sub- and
superadiabatic cases look similar except that they have,
respectively, everywhere positive and negative $N_0^2$ that diverges
at the surface because the sound speed vanishes there.

\subsection{Perturbation equations}

We consider here small axisymmetric ($\partial/\partial y=0$)
perturbations of the form ${\bf u'}, \rho', p', \psi' \propto
exp(-i\omega t+ikx)$ about the equilibrium states found above, where
$\omega$ and $k$ are the frequency and the radial wavenumber,
respectively. Without a loss of generality, we assume throughout
that $k$ is non-negative, $k\geq 0$. After switching to
non-dimensional variables:
\[
t \rightarrow \Omega t,~~\omega\rightarrow
\frac{\omega}{\Omega},
\]
\[
k\rightarrow \frac{kc_{sm}}{\Omega},~~N_0\rightarrow
\frac{N_0}{\Omega},~~g_0\rightarrow \frac{g_0}{\Omega c_{sm}},
\]
\[
u'_x\rightarrow \frac{u'_x}{c_{sm}},~~~u'_y\rightarrow
\frac{u'_y}{c_{sm}},~~~u'_z\rightarrow \frac{u'_z}{c_{sm}},
\]
\[
\rho'\rightarrow \frac{\rho'}{\rho_m},~~~p'\rightarrow
\frac{p'}{\rho_m c_{sm}^2},~~~\psi'\rightarrow
\frac{\psi'}{c_{sm}^2},
\]
we can derive from equations (1-4) the following set of equations
governing the linear dynamics of axisymmetric perturbations
\begin{equation}
-i\omega u'_x=-\frac{ikp'}{\rho_0}+2u'_y-ik \psi'
\end{equation}
\begin{equation}
-i\omega u'_y=(q-2)u'_x
\end{equation}
\begin{equation}
-i\omega
u'_z=-\frac{1}{\rho_0}\frac{dp'}{dz}-g_0\frac{\rho'}{\rho_0}-\frac{d\psi'}{dz}
\end{equation}
\begin{equation}
-i\omega \rho'+ik\rho_0u'_x+\frac{d}{dz}(\rho_0u'_z)=0
\end{equation}
\begin{equation}
-i\omega(p'-c_s^2\rho')+\frac{\rho_0c_s^2N_0^2}{g_0}u'_z=0
\end{equation}
\begin{equation} \left(-k^2+\frac{d^2}{dz^2}\right)\psi'=\frac{\rho'}{Q}.
\end{equation}
If we now make the changes: $iu'_z\rightarrow u'_z,~\omega
p'/\rho_0\rightarrow p',~\omega\psi'\rightarrow \psi'$ and eliminate
$u'_x,~u'_y,~\rho'$ from equations (10), (11), (13), we arrive at
the following set of equations for the three basic quantities
$u'_z,~p',~\psi'$ (henceforth primes will be omitted)
\begin{equation}
\frac{du_z}{dz}=\frac{g_0}{c_s^2}u_z-\left(\frac{1}{c_{s}^2}-
\frac{k^2}{\omega^2-\kappa^2}\right)p+\frac{k^2\psi}{\omega^2-\kappa^2}
\end{equation}
\begin{equation}
\frac{dp}{dz}=\frac{N_0^2}{g_0}p+(\omega^2-N_0^2)u_z-\frac{d\psi}{dz}
\end{equation}
\begin{equation}
\frac{d^2\psi}{dz^2}-k^2\psi=\frac{\rho_0}{Q}\left(\frac{p}{c_s^2}+\frac{N_0^2}{g_0}u_z\right),
\end{equation}
where the non-dimensional epicyclic frequency $\kappa$ is given by
$\kappa^2=2(2-q)=1$. Notice that the density perturbations on the
right hand side of the linearized Poisson's equation (18) consist of
two physically different parts. The density perturbations due to
compressibility, which are inversely proportional to the squared
sound speed and proportional to the pressure perturbation, and the
density perturbations due to the stratified background, which are
proportional to the vertical velocity perturbation and $N_0^2$. This
latter term due to stratification is obviously absent in the
two-dimensional analysis of gravitational instability as well as
when the disc is adiabatic (see GLB). Equations (16-18) form the
basis for determining the axisymmetric normal modes of perturbations
existing in compressible, stratified, polytropic, self-gravitating
discs. In the non-self-gravitating limit, these equations reduce to
the main equations of KP. The reflection symmetry of the unperturbed
disc allows us to take the eigenfunctions/modes to be either even or
odd with respect to $z$. So, to determine the eigenfrequencies
(dispersion diagrams, that is, $\omega$ as a function of $k$ for
various branches) and eigenfunctions, we can numerically integrate
the main equations (16-18) only in the upper half, $0\leq z \leq h$,
of the full vertical extent of the disc provided that suitable
boundary conditions are imposed at $z=0$ and $z=h$. In other words,
we need to solve a boundary value problem. Here the disc height $h$,
as noted above, is determined by the parameters $Q$ and $s$. By
inspecting equations (16-18), it is clear that a normal mode whose
pressure $p$ and potential $\psi$ perturbations are even functions
under reflection in $z$, has odd vertical velocity $u_z$ and
gravitational potential derivative $d\psi/dz$ perturbations under
reflection and vice versa. We define a normal mode as being `even'
(`odd') if the corresponding vertical velocity and gravitational
potential derivative are odd (even) functions of $z$ \citep[this
convention agrees with that of][but differs from LP and KP]{Og98}.
Thus, the boundary conditions at the midplane, $z=0$, for the even
modes are:
\begin{equation}
u_z(0)=0,~d\psi/dz(0)=0
\end{equation}
and for the odd modes are:
\begin{equation}
p(0)=0,~\psi(0)=0.
\end{equation}
At $z=h$ we impose the usual free-surface boundary condition for
which the Lagrangian pressure perturbation vanishes. In our new
non-dimensional variables this translates as
\begin{equation}
p=g_0u_z~~~~at~~~~z=h.
\end{equation}
The boundary condition for the gravitational potential can be
derived by treating the surface displacement as a surface
distribution of gravitating matter at $z=h$. Using the continuity of
potential across the boundary, Gauss's flux theorem and the fact
that outside the disc, at $z\rightarrow \pm \infty$, gravitational
potential should vanish, we arrive at the following condition at the
disc surface \citep[see e.g., GLB,][]{LP93b}
\begin{equation}
\frac{d\psi}{dz}+k\psi=-\frac{u_z\rho_0}{Q}~~~~at~~~~z=h.
\end{equation}
Equations (16-18) supplemented with boundary conditions (21),(22) at
the surface $z=h$ and (19),(20) at the midplane fully determine a
boundary value problem. Using these boundary conditions and
variational principle, it can be easily shown that $\omega^2$ is
real (GLB, RPL) that much alleviates the search of eigenfrequencies.
We integrate these equations from $z=h$ downwards to $z=0$
separately for the even modes under conditions (19) and for the odd
modes under conditions (20). We will see below that in the presence
of self-gravity, the dispersion diagrams for these two mode parities
are quite different. In other words, self-gravity influences even
and odd modes, or changes their dispersion characteristics, in
different ways. As is typical, for a given equilibrium structure,
that is, for given $Q$ and $s$, and for a given radial wavenumber
$k$, midplane boundary conditions can be satisfied only for certain
values of $\omega$, yielding dispersion relations $\omega(k)$ for
different branches of modes classified below. We use a Runge-Kutta
integrator and root-finding algorithms of MATLAB package to first
numerically solve differential equations (16-18) and then find
eigenfrequencies.

\section{The classification of vertical modes}

In this section, we for the moment remove self-gravity (put $\psi
\rightarrow 0, Q\rightarrow \infty$) from the perturbation equations
(16-18) and from the equilibrium in order to classify and
characterize all the vertical axisymmetric normal modes existing in
a three-dimensional disc. In special cases, an analogous
classification of modes in non-self-gravitating discs has been done
previously by several authors \citep[RPL, LP, KP,][]{Og98}. The aim
here is to briefly review and synthesize the results from these
studies. In the next section, we will again switch on self-gravity
and investigate how it affects the characteristics of these modes.
Accordingly, in the non-self-gravitating case, we need the midplane
and surface boundary conditions discussed above, but only for
pressure and vertical velocity perturbations.

Figure 2 shows the typical dispersion diagrams for three types of
equilibria: subadiabatic ($1+1/s<\gamma$), adiabatic
($1+1/s=\gamma$) and superadiabatic ($1+1/s>\gamma$)
stratifications. In the subadiabatic case, one can identify four
distinct classes of vertical normal modes \citep[see also
KP,][]{Og98}. These modes are: the acoustic p-modes, the surface
gravity f-modes, the buoyancy g-modes and the inertial r-modes the
restoring forces for which at large radial wavenumbers $kh\gg 1$ are
mainly provided, respectively, by compressibility/pressure, by the
displacements of free surface of the disc, by buoyancy due to the
vertical entropy gradient and by inertial forces due to disc
rotation. The existence of the g-modes in polytropic discs is
attributable to the variation of the sound speed and $N_0$ with
height; the latter diverges at the disc surface giving rise to the
trapped g-mode there. For $1+1/s<\gamma$, all these modes have
$\omega^2>0$ and, therefore, are stable. The p-, f- and g-modes are
high-frequency modes with frequencies always larger than the
epicyclic frequency, $\omega^2\geq\kappa^2$, while the r-mode is of
low-frequency with $\omega^2\leq\kappa^2$ (we remind that hereafter
$\kappa=1$).
\begin{figure*}
\includegraphics[width=0.33\textwidth, height=0.28\textwidth]{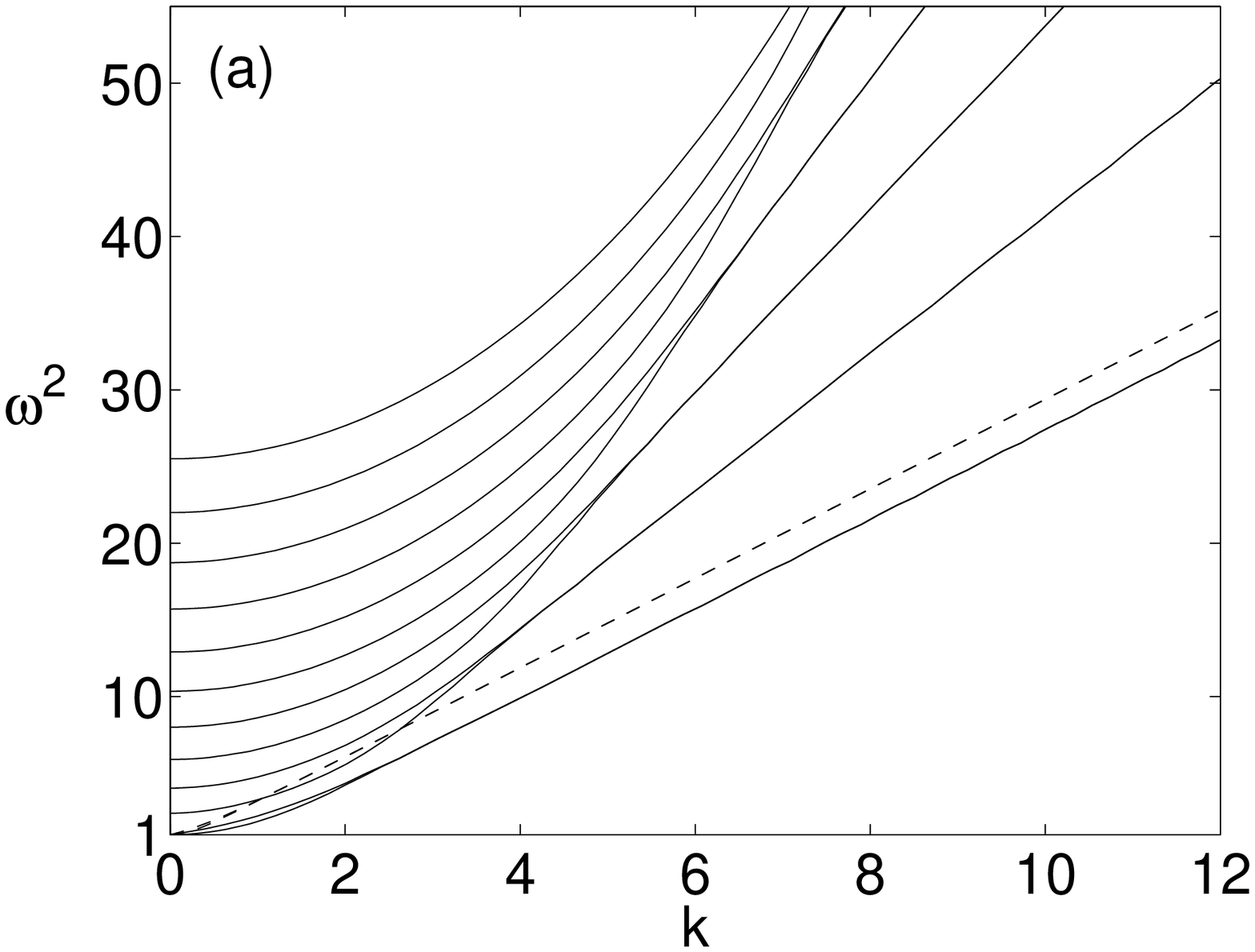}
\includegraphics[width=0.33\textwidth, height=0.28\textwidth]{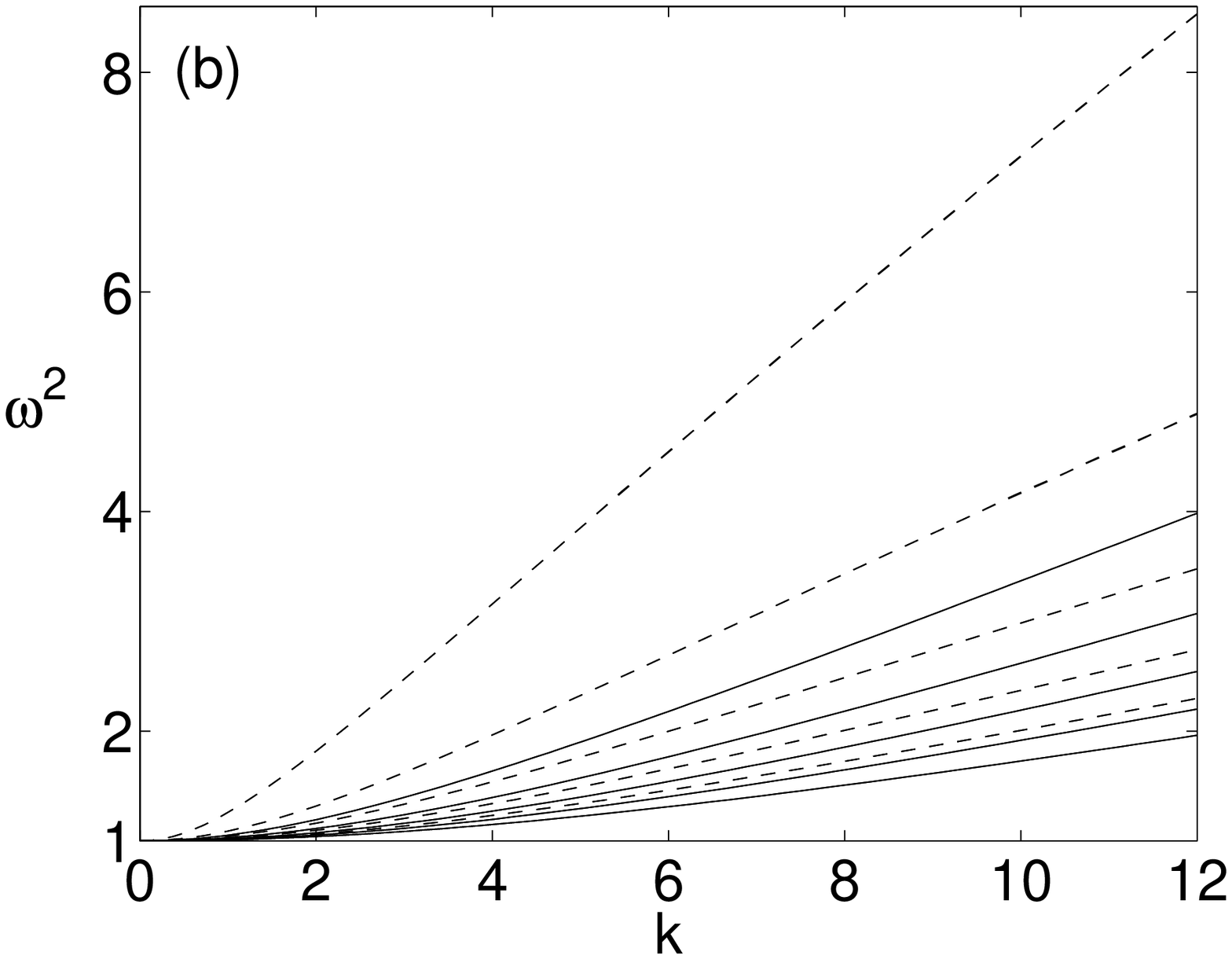}
\includegraphics[width=0.33\textwidth, height=0.28\textwidth]{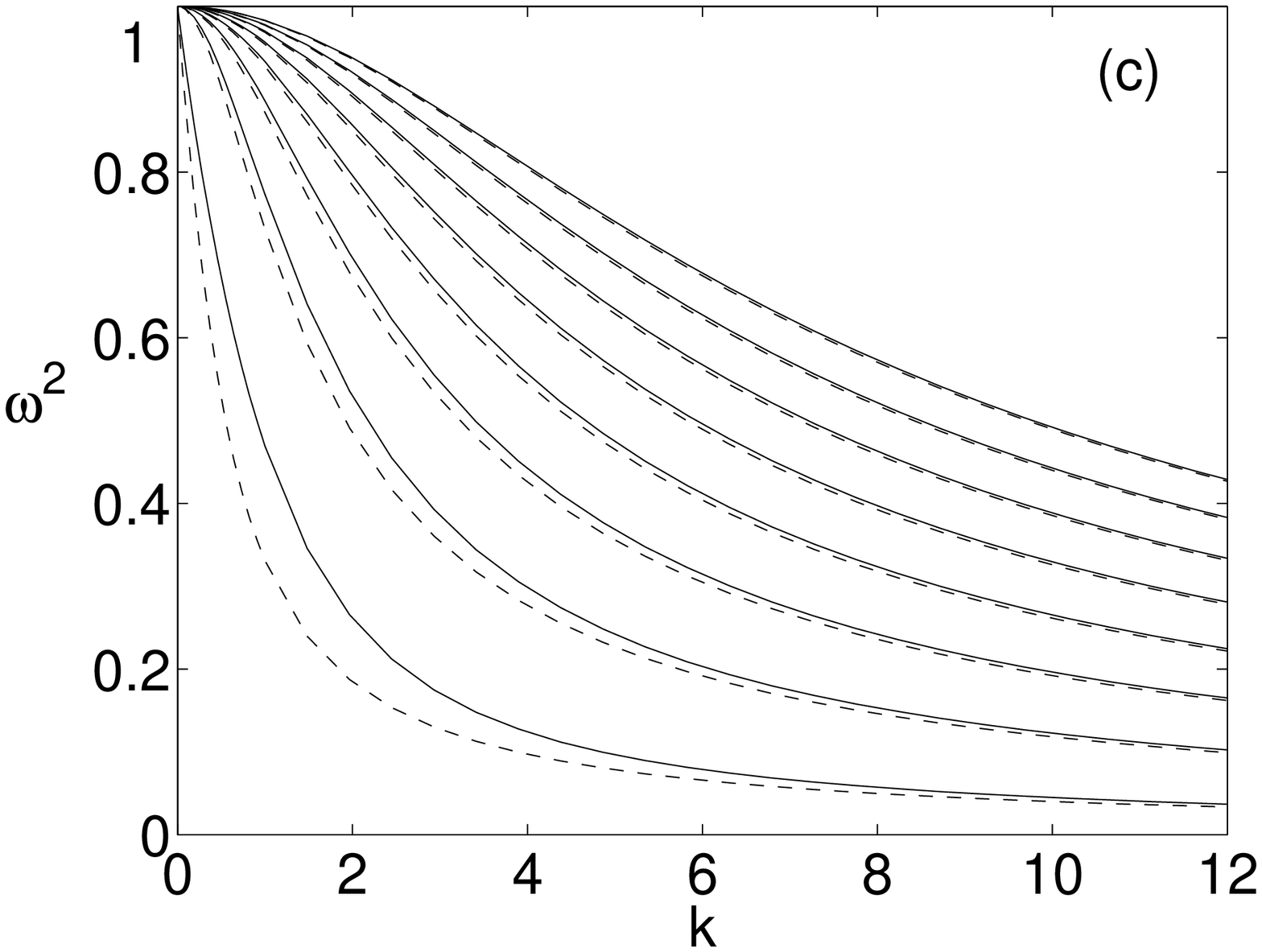}
\includegraphics[width=0.335\textwidth, height=0.28\textwidth]{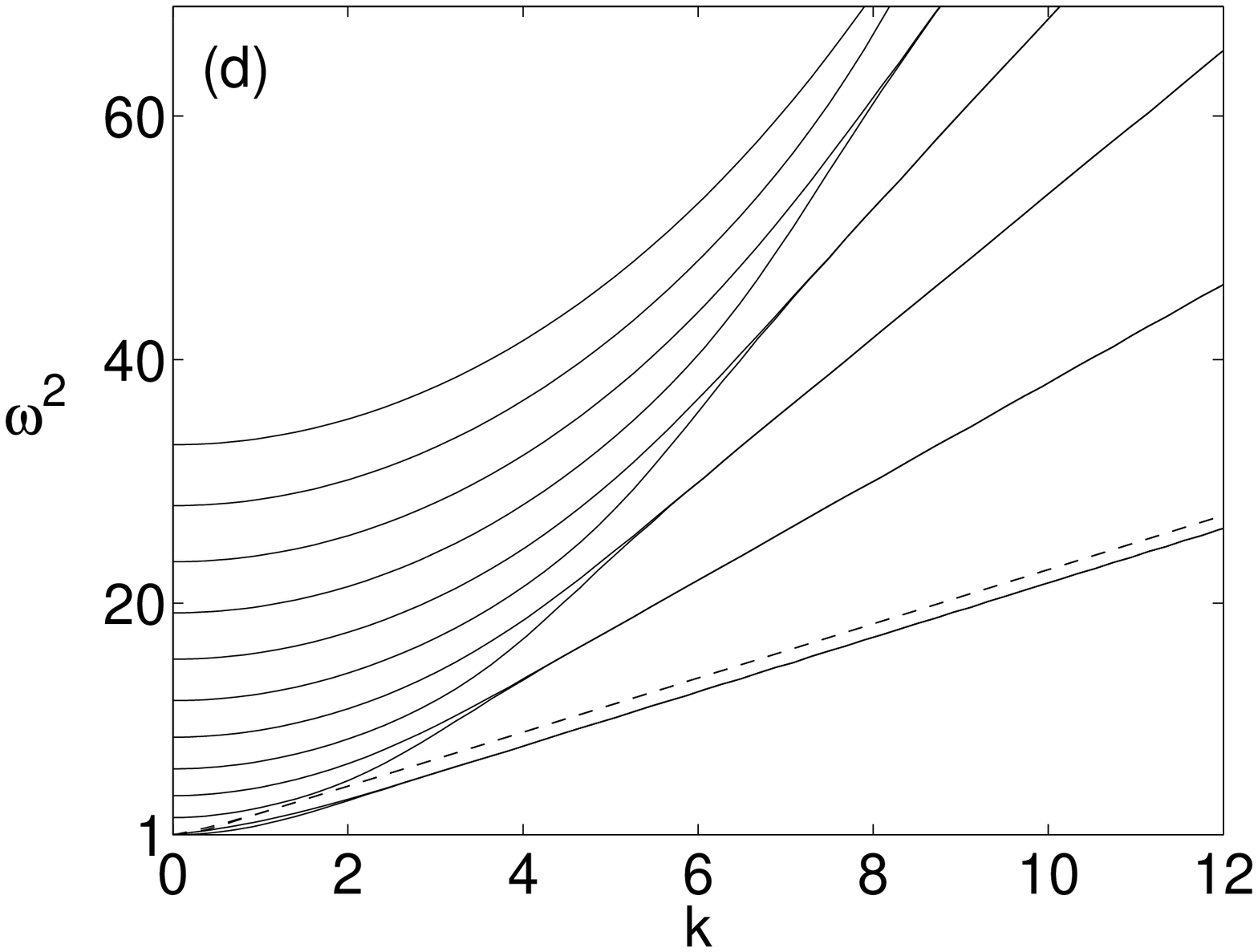}
\includegraphics[width=0.335\textwidth, height=0.28\textwidth]{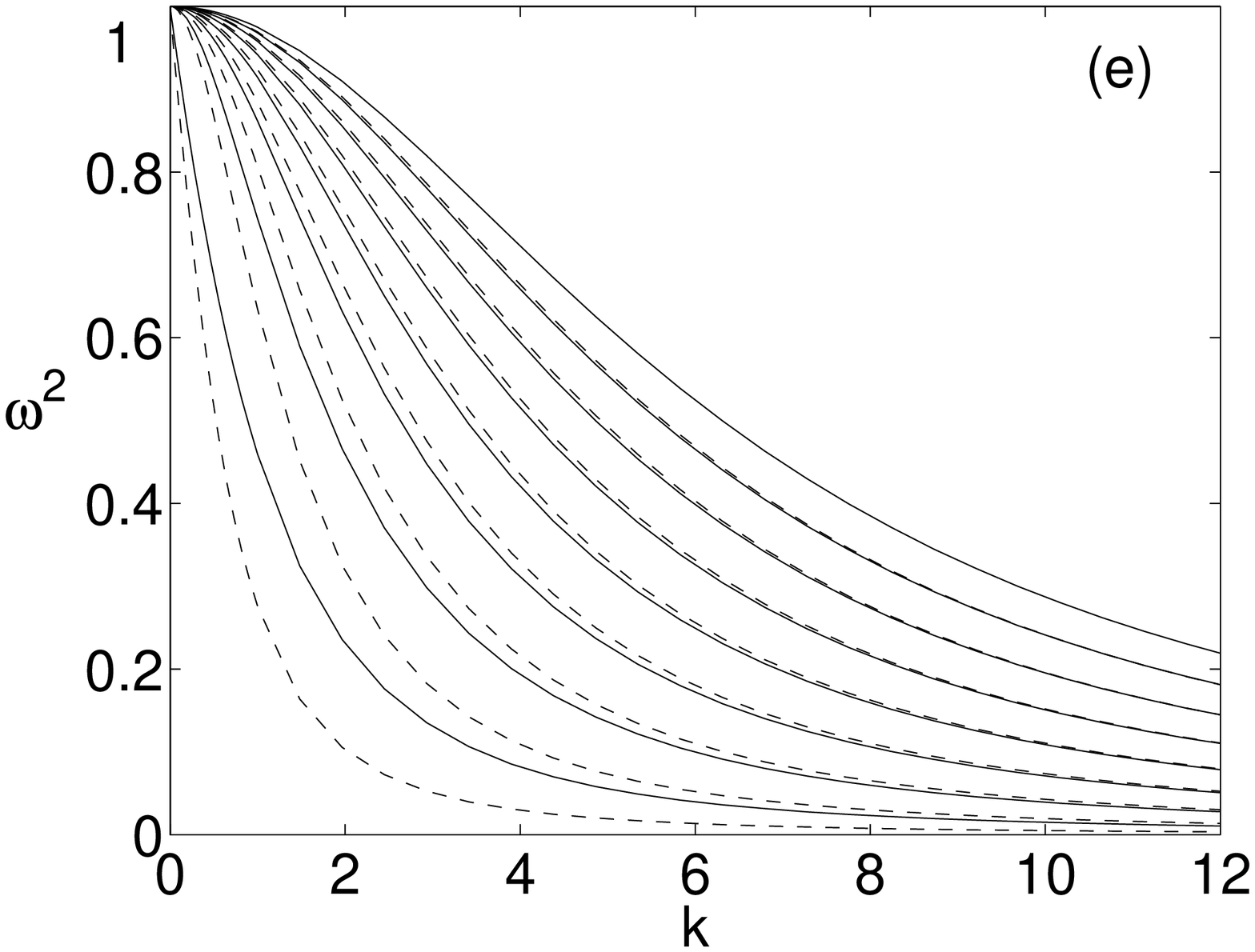}
\includegraphics[width=0.335\textwidth, height=0.28\textwidth]{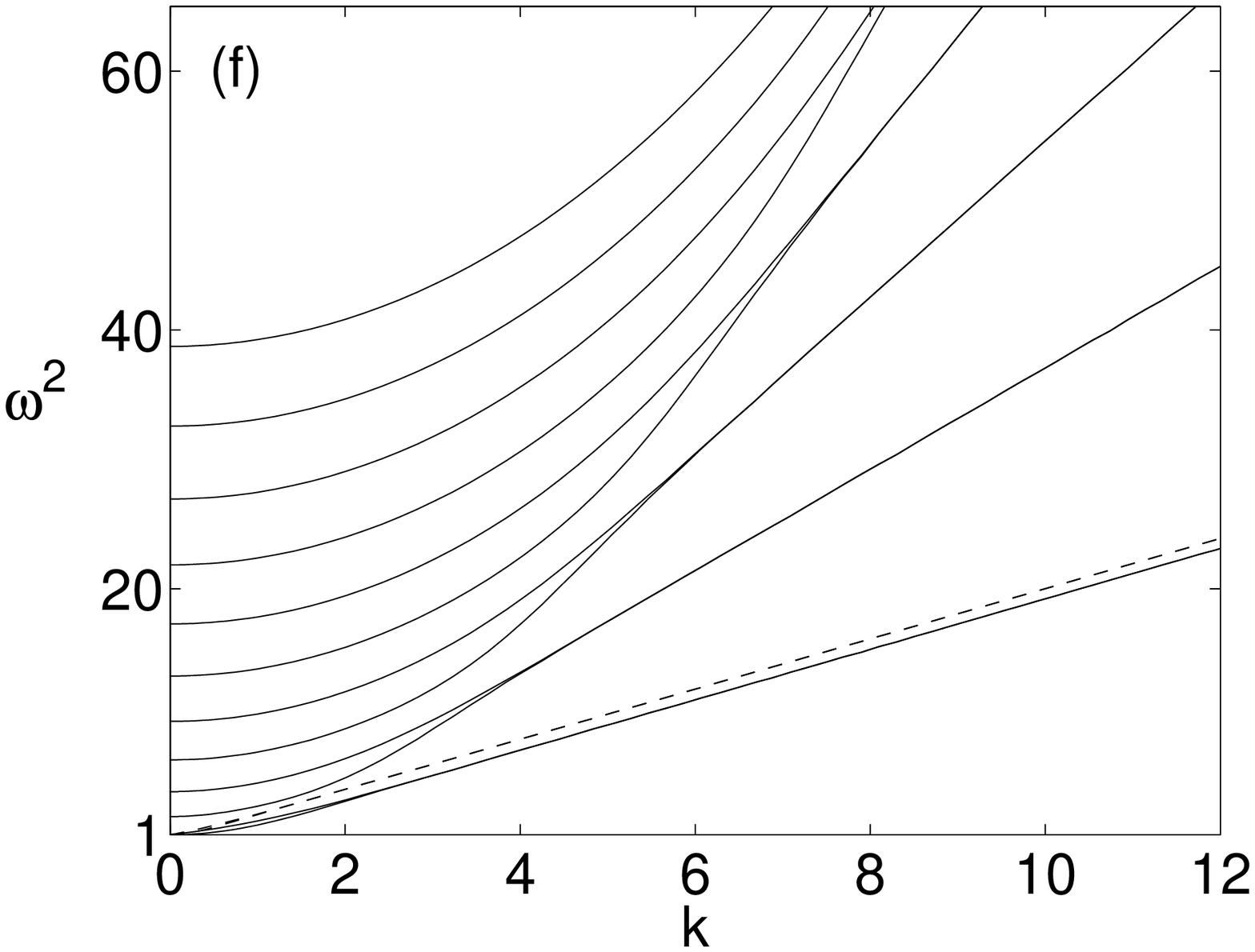}
\includegraphics[width=0.335\textwidth, height=0.28\textwidth]{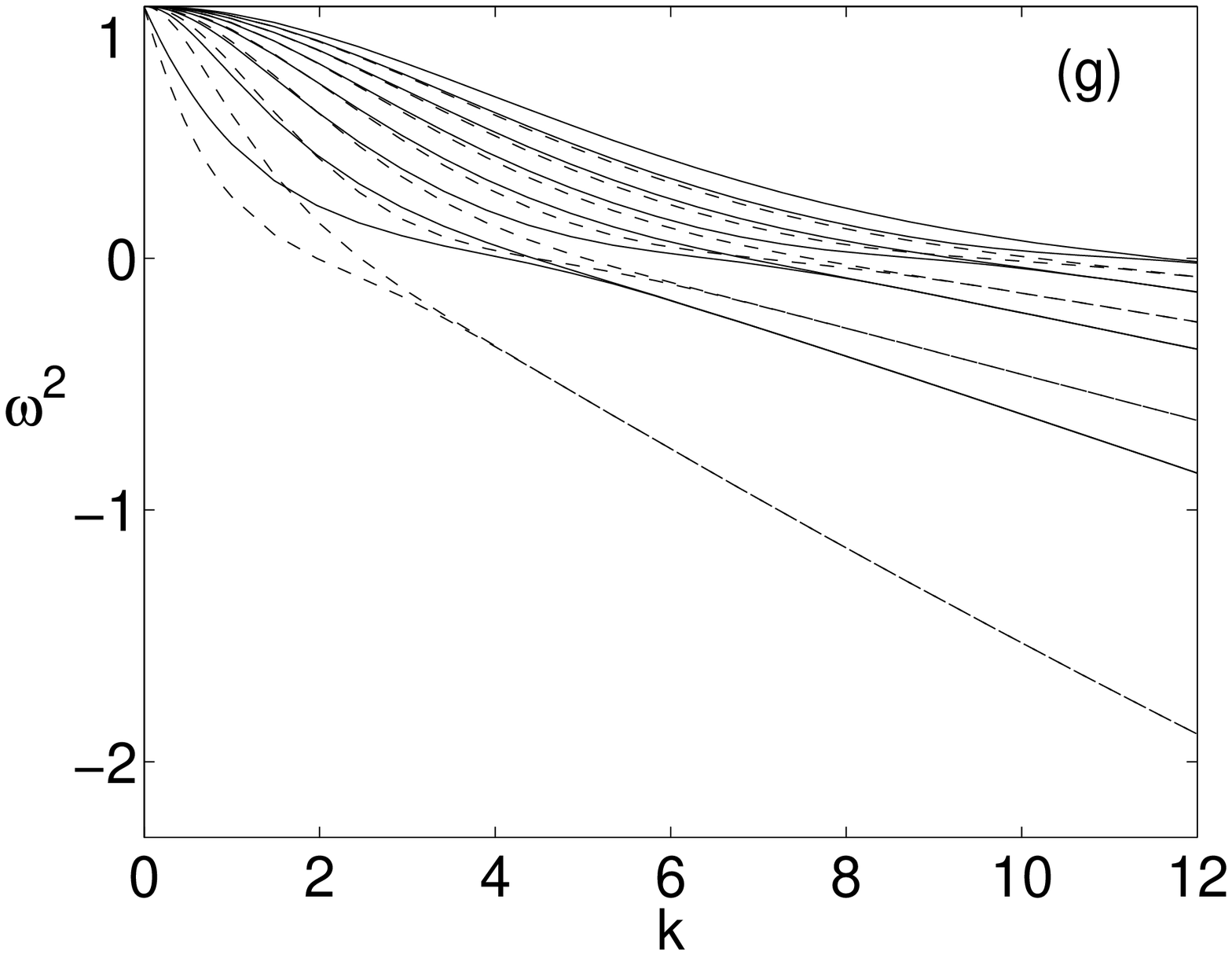}
\caption{Dispersion diagrams in the non-self-gravitating case for a
subadiabatic disc with $s=5$ (panels a, b, c), for an adiabatic disc
with $s=2.5$ (panels d, e) and for a superadiabatic disc with
$s=1.7$ (panels f, g). The solid lines in panels (a),(d),(f) show
the frequencies for different branches of the surface gravity f- and
acoustic p-modes vs. radial wavenumber $k$. In order of increasing
frequency these branches/curves are ${\rm
f^{e},f^{o},p_{1}^{e},p_{1}^{o},p_{2}^{e},p_{2}^{o},
p_{3}^{e},p_{3}^{o}, p_{4}^{e},p_{4}^{o}, p_{5}^{e},p_{5}^{o}}$
(superscripts denote the even and odd modes, which are numbered,
respectively, according to the number of nodes of vertical velocity
and pressure perturbations in the interval $0 < z\leq h$). Although,
for large $k$, the frequencies of even and odd modes coalesce. The
dashed lines in these panels correspond to frequencies of the
f-modes computed in the incompressible limit. The solid lines in
panel (b) show the convectively stable g-mode branches, which in
order of decreasing frequency are ${\rm
g_1^{o,e},g_2^{o,e},g_3^{o,e},g_4^{o,e},g_5^{o,e}}$. Here the
frequencies of even and odd modes coincide. The solid lines in
panels (c),(e) show the r-mode branches, which in order of
increasing frequency are ${\rm
r_{0}^{o},r_{0}^{e},r_{1}^{o},r_{1}^{e}, r_{2}^{o},r_{2}^{e},
r_{3}^{o},r_{3}^{e}}$ \citep[${\rm r_{0}^{e}}$ and ${\rm r_{0}^{o}}$
are the basic even and odd r-modes and subscript `0' means that,
respectively, vertical velocity and pressure perturbations for them
have no nodes in the interval $0 < z\leq h$. In this respect, our
numbering of modes differs from that of][]{Og98}. The solid lines in
panel (g) show the convectively unstable g-modes, which in order of
increasing $\omega^2$ are ${\rm g_{1}^{o}, g_{0}^{e}, g_{2}^{o},
g_{1}^{e}, g_{3}^{o}, g_{2}^{e}, g_{4}^{o}, g_{3}^{e}}$ (${\rm
g_{0}^{e}}$ denotes the basic even g-mode, the corresponding
vertical velocity perturbation for which has no nodes in the range
$0 < z\leq h$). Similar to the p-modes, the convectively unstable
even and odd g-modes coalesce at large $k$. In each panel, the
dashed lines show the corresponding mode branches computed for the
incompressible case with the same ordering of eigenfrequencies. From
panels (b),(g), we see that compressibility most strongly affects
the convectively stable and unstable g-modes.}
\end{figure*}

For adiabatic stratification, there is no buoyancy $(N_0^2=0)$ and,
therefore the g-mode disappears, while other modes remain
qualitatively unchanged. Similarly, the r-mode does not exist in a
disc that has zero epicyclic frequency. For $1+1/s>\gamma$, the
g-mode becomes convectively unstable ($\omega^2<0$) for radial
wavenumbers larger than a certain value because of negative
buoyancy. Its characteristic timescale (growth rate) can be of the
order of the epicyclic frequency or less and due to this, it
interferes with motions corresponding to the r-mode. Consequently,
in the superadiabatic case, the r- and g-modes merge at
$\omega^2\leq \kappa^2$ and appear as a single mode, which we still
call the g-mode and not the r-mode, because the behaviour of
corresponding eigenfunctions with height in this case is similar to
that of the convectively stable g-mode (see also RPL); the p- and
f-modes are not much affected.

All these modes come in even and odd pairs. In Fig. 2, even (odd)
modes are numbered according to the number of nodes of the vertical
velocity (pressure) perturbation in the interval $0 < z\leq h$ (the
node at $z=0$ is not counted). We also see the coalescence of the
dispersion curves of the even and odd p- and f-modes and also the
convectively unstable even and odd g-modes as $k$ increases past a
certain point, which depends on the mode number (RPL, KP). This is
associated with a transition from oscillatory to evanescent
behaviour of eigenfunctions. We do not plot eigenfunctions here, as
an extensive discussion of their properties can be found in RPL, KP,
LO98 and \citet{Og98}. We only mention that the eigenfunctions of
the p- and g-modes are trapped near the surfaces and decay towards
the midplane, while the eigenfunctions of the r-mode are
concentrated near the midplane and decay towards the surfaces. As
for the eigenfunctions of the fundamental f-mode, they have no nodes
and monotonically decay from the surfaces to the midplane. Below we
show that these properties of eigenfunctions are altered due to
self-gravity. Specifically, the number of nodes along each branch of
the dispersion diagrams, which is preserved in the
non-self-gravitating limit, is no longer constant in the presence of
self-gravity. The frequencies of the p- and r-modes increase with
mode number. The frequencies and growth rates of the convectively
stable and unstable g-modes, respectively, decrease with the mode
number. These are the well-known general properties of vertical
modes in polytropic discs.

\begin{figure*}
\includegraphics[width=0.9\textwidth, height=0.7\textwidth]{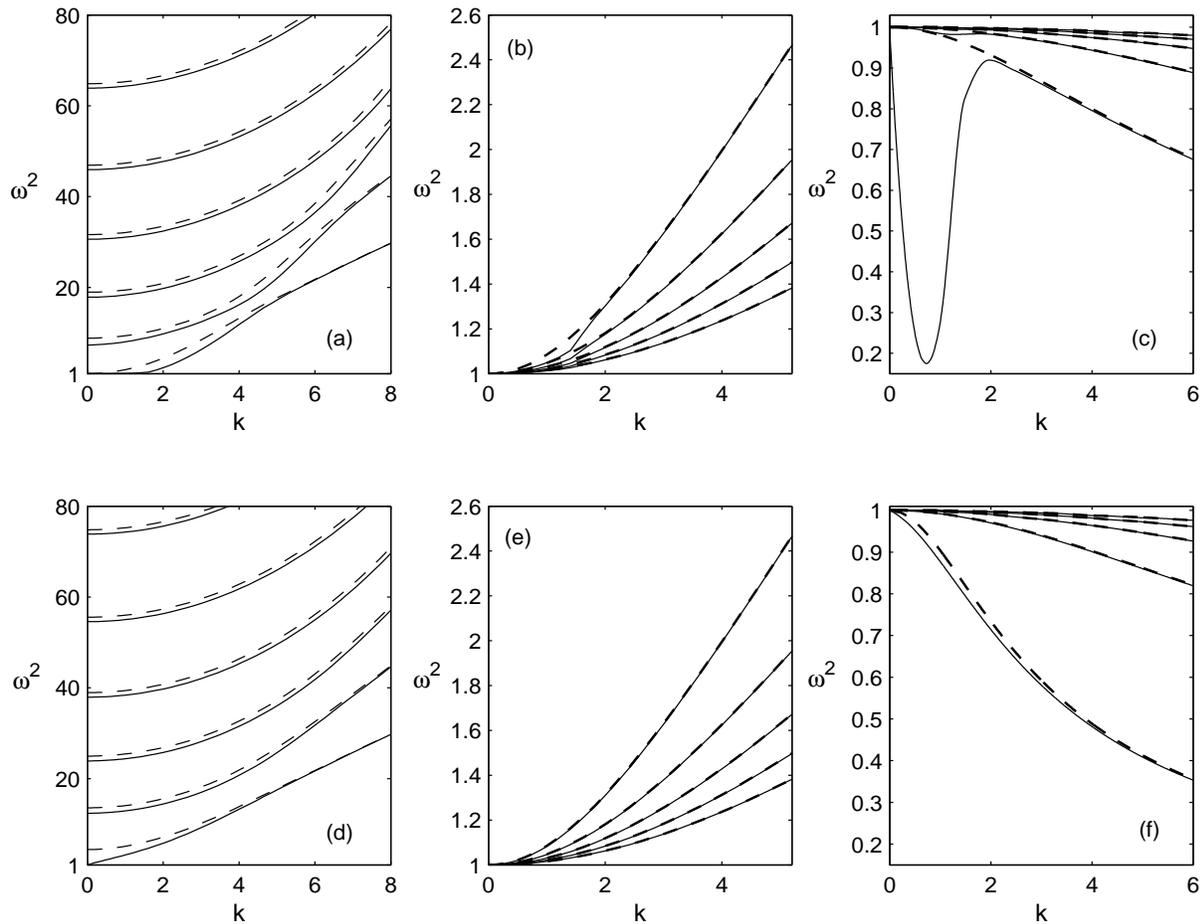}
\caption{Dispersion diagrams of vertical modes in a subadiabatic
self-gravitating disc with $s=5$ and $Q=0.2$. Shown are (a) the even
p- and f-modes, (b) the even g-mode, (c) the even r-mode, (d) the
odd p- and f-modes, (e) the odd g-mode, (f) the odd r-mode. Dashed
lines are the corresponding branches computed without self-gravity
in the perturbations, but with the same self-gravitating
equilibrium. The frequency ordering and mode naming are the same as
in Fig. 2. Self-gravity reduces/shifts the frequencies of the even
and odd p-, f-modes and the even g- and r-modes mainly in the range
$0 \leq k \leq 2$; the frequencies of the odd g- and r-modes are
almost unaffected by self-gravity. The frequency of the basic even
r-mode (denoted above as ${\rm r_{0}^{e}}$) is modified most
strongly by self-gravity as evidenced by the largest dip on the
corresponding dispersion curve in panel (c). For large $k$ and/or
large mode numbers, the effect of self-gravity becomes weak and the
dispersion curves merge with the dashed ones for the
non-self-gravitating case.}
\end{figure*}

Although each mode type has its dominant restoring force, one out of
the above mentioned four types (compressibility, surface gravity,
buoyancy, inertial forces), the other three forces also contribute
to a certain degree for small radial wavenumbers $kh \lsim 1$. For
example, the p- and f-modes are modified both by rotation/inertial
forces and buoyancy, the f-mode is also modified by compressibility,
the g-mode is modified by rotation and compressibility and the
r-mode is modified by buoyancy and compressibility. For the sake of
comparison with the relevant result from previous studies, we are
particularly interested in to what degree compressibility modifies
generally non-compressive f-, g- and r-modes. So, we decided to
juxtapose the dispersion diagrams for these modes computed
separately in the compressible and incompressible cases. We take the
incompressible limit by formally letting the adiabatic index go to
infinity, $\gamma\rightarrow \infty$ \citep{Og98}. After that
equations (16-18) without self-gravity take the form
\[
\frac{du_z}{dz}=\frac{k^2}{\omega^2-\kappa^2}p \nonumber
\]
\[
\frac{dp}{dz}=\frac{N_0^2}{g_0}p+(\omega^2-N_0^2)u_z. \nonumber
\]
We solved these equations with the same boundary conditions and
found the dispersion diagrams of the f-, g- and r-modes that survive
in this limit (obviously, the acoustic p-mode disappears). The
results are plotted in Fig. 2 with dashed lines. Notice that by
taking the incompressible limit in this manner, we have been able to
retain the f-mode, as expected. Instead, KP set the density
perturbations to zero in the linearized continuity equation
(anelastic approximation) that resulted in the f-mode disappearing
in their incompressible limit. We see that compressibility most
strongly affects the convectively stable and unstable g-mode
branches with small mode numbers (Figs. 2b,2g), or equivalently with
vertical extent comparable to the disc height, even at large radial
wavenumbers \citep{Og98}. The frequencies of the f- and r-modes do
not change much, indicating that these modes are nearly
incompressible.

\section{vertical modes in the presence of self-gravity}

Having characterized all types of axisymmetric normal modes in a
stratified disc, let us now compute the dispersion diagrams taking
into account self-gravity in the perturbation equations and in the
equilibrium. This will allow us to understand how the frequencies
and the structure of the eigenfunctions of the above-described mode
types are altered by self-gravity, which mode acquires the largest
positive growth rate in the presence of self-gravity and, thus
determines the onset criterion and nature of gravitational
instability of a disc. In other words, we return to the boundary
value problem formulated in Section 2, which is represented by
equations (16-18) supplemented with boundary conditions (21),(22) at
the surface and conditions (19),(20) of the even and odd symmetry of
a solution at the midplane. For each $Q$ and $s$, we first determine
the corresponding vertical distribution of the equilibrium
quantities in equations (16-18), $\rho_0, c_s^2, g_0, N_0^2$, with
height as described in Section 2 and then based on this compute the
normal modes.

Figures 3,4,5 and 6 show the typical dispersion diagrams for the p-,
f-, g- and r-modes in a self-gravitating disc for subadiabatic,
adiabatic and superadiabatic vertical stratifications. We separately
plot the dispersion diagrams of the even and odd parities for each
mode type to clearly see the influence of self-gravity on them,
which, as evident from these figures, depends on the mode parity.
Unlike in the non-self-gravitating limit, the number of nodes of the
vertical velocity and pressure perturbations in the presence of
self-gravity are not preserved along each mode branch. However, as
we will see below, at large $k$, the influence of self-gravity on
the mode dynamics is small and the dispersion diagrams merge with
their non-self-gravitating counterparts (shown with dashed curves in
Figs. 3, 5). So, the naming of the modes for smaller $k$, where the
effect of self-gravity is important, is done by continuity with the
large-$k$ limit for each mode branch.

\begin{figure}
\centering\includegraphics[width=\columnwidth]{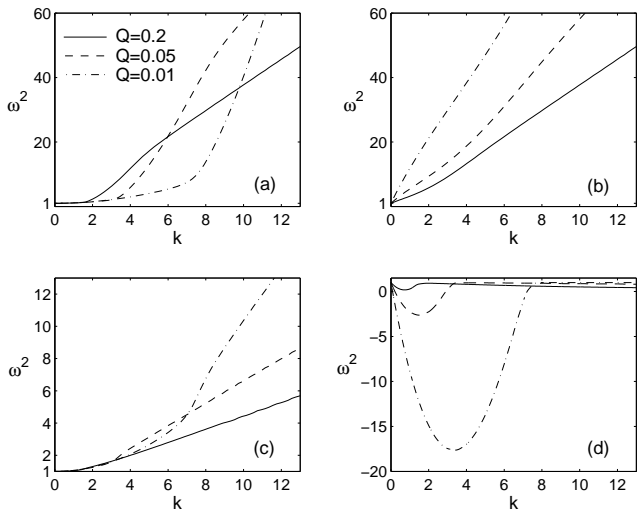}
\caption{Dispersion curves for $s=5$ and $Q=0.01$ (dashed-dotted
lines), $Q=0.05$ (dashed lines), $Q=0.2$ (solid lines). Panels
(a),(b) show the even and odd f-modes, respectively. Panel (c) shows
the basic even g-mode and panel (d) shows the basic even r-mode. The
frequencies of the f- and g-modes never fall below the epicyclic
frequency with decreasing $Q$ and, therefore, are always stable. The
basic even r-mode, which is most significantly affected by
self-gravity, becomes unstable when the dip on the corresponding
dispersion curve crosses the $\omega^2=0-$axis. This dip deepens and
broadens with decreasing $Q$.}
\end{figure}

\subsection{Subadiabatic stratification}

Consider first the subadiabatic case (Fig. 3), where we again have
the p-, f-, g- and r-modes, but their dispersion diagrams are
modified/shifted from their non-self-gravitating counterparts
towards lower values due to self-gravity. As illustrated in Figs.
3a,3d, self-gravity reduces the frequencies of all branches of the
p- and f-modes, for both even and odd parities, but it more affects
the f-modes. The situation for the g- and r-modes is different
(Figs. 3b,3c,3e,3f): only the frequencies of the even g- and r-modes
are reduced by self-gravity mostly for radial wavenumbers in the
range $0\leq k \leq 2$ (dips on the corresponding dispersion curves
in Figs. 3b,3c indicating deviations from the non-self-gravitating
dashed ones), whereas the frequencies of the odd g- and r-modes are
almost unaffected by self-gravity. The frequencies of the
fundamental f-modes, the first few branches of the p-modes and also
the first few branches of the even g- and r-modes are reduced
noticeably. With increasing mode number and/or radial wavenumber,
the effect of self-gravity on the eigenfrequencies gradually falls
off, because the corresponding eigenfunctions become of smaller and
smaller horizontal and/or vertical scale (the number of nodes
increases) and take the form similar to that in the
non-self-gravitating limit. As a result, the dispersion diagrams
become more and more identical to their non-self-gravitating
counterparts. The ordering of frequencies for the even and odd p-,
f-, g- and r-modes, as it is in the non-self-gravitating case (see
Fig. 2): $\omega^2({\rm r})\leq \kappa^2 \leq \omega^2({\rm
g})<\omega^2({\rm f})<\omega^2({\rm p})$, is preserved in the
self-gravitating case as well, however small the value of $Q$. Note
also in Fig. 4 that the frequencies of the the even and odd f-modes
and also of the basic even g-mode, for which the influence of
self-gravity is stronger, never fall below the epicyclic frequency
with decreasing $Q$ and, therefore, these modes always remain
gravitationally stable (we also checked that the same situation
holds for the dispersion curves of the even and odd f-modes in the
adiabatic and superadiabatic cases described below). This is a
consequence of the three-dimensionality of the disc. Thus, as
evident from Figs. 3c,4d, self-gravity most significantly affects
the basic branch of the even r-mode and only this branch can become
unstable due to self-gravity. From Fig. 4d, it is seen that the dip
on this branch deepens and broadens with decreasing $Q$. (We denote
the wavenumber at which the minimum of this dip is located by
$k_m$.) The gravitational instability sets in when the dip's minimum
first touches the $\omega^2=0-$axis at some $Q_{cr}$ and then
farther extends into the unstable $\omega^2<0$ region for
$Q<Q_{cr}$. We will demonstrate in Section 5 that the basic even
r-mode becomes strongly compressible at $k\sim k_m$ (Fig. 8), as
opposed to the r-mode in a non-self-gravitating disc, and the
density perturbations due to compressibility are responsible for its
gravitational instability. For $s=5$, we find $Q_{cr}=0.168$, which
gives $H=2h=2.96$ for the disc total thickness, and $k_{m}=0.8$ (see
Fig. 12). Note that the characteristic radial scale of instability
$\lambda_{m}=2\pi/k_{m}=7.85$ is not much greater than the disc
thickness $H$. In self-gravitating discs, compressibility and
inertial forces play an important role in the dynamics of the basic
even r-mode at wavelengths $\lambda \sim \lambda_m \gsim H$ and so
it resembles a 2D density wave at such wavelengths (see Section
4.4). The fact that the characteristic scale of gravitational
instability $\lambda_m$ turns out to comparable to the disc
thickness may offer a clue why angular momentum transport due to
self-gravity tends to be a local phenomenon
\citep[e.g.,][]{Gam01,LR04,Betal06}. In this respect, it should also
be mentioned that the analogous gravitational instability of
low-frequency modes in a strongly compressed gaseous slab was also
found by \citet{LP93b}. In their analysis, these modes, called
neutral modes, are basically degenerate r-modes, since the
compressed gaseous slab was not rotating.

\subsection{Adiabatic stratification}

In the adiabatic case (Figs. 5a-5d), there is no g-mode and the
behaviour of the p-, f- and r-modes under self-gravity has
qualitatively the same character as in the subadiabatic case above.
Again, the basic branch of the even r-mode appears to be most
significantly affected by self-gravity (the corresponding dispersion
curve in Fig. 5c has the same form as that in Fig. 3c) and becomes
gravitationally unstable. The $\omega^2$ of the p- and f-modes,
although lowered by self-gravity, always remain larger or equal to
$\kappa^2$ irrespective of the value of $Q$. In this case, $s=2.5$
and the corresponding $Q_{cr}=0.177$ and $k_{m}=0.81$. Again the
radial scale of instability $\lambda_{m}=2\pi/k_{m}=7.75$ is not far
from the disc thickness for these parameters, $H=2h=2.08$.

\begin{figure*}
\includegraphics[width=\textwidth, height=0.7\textwidth]{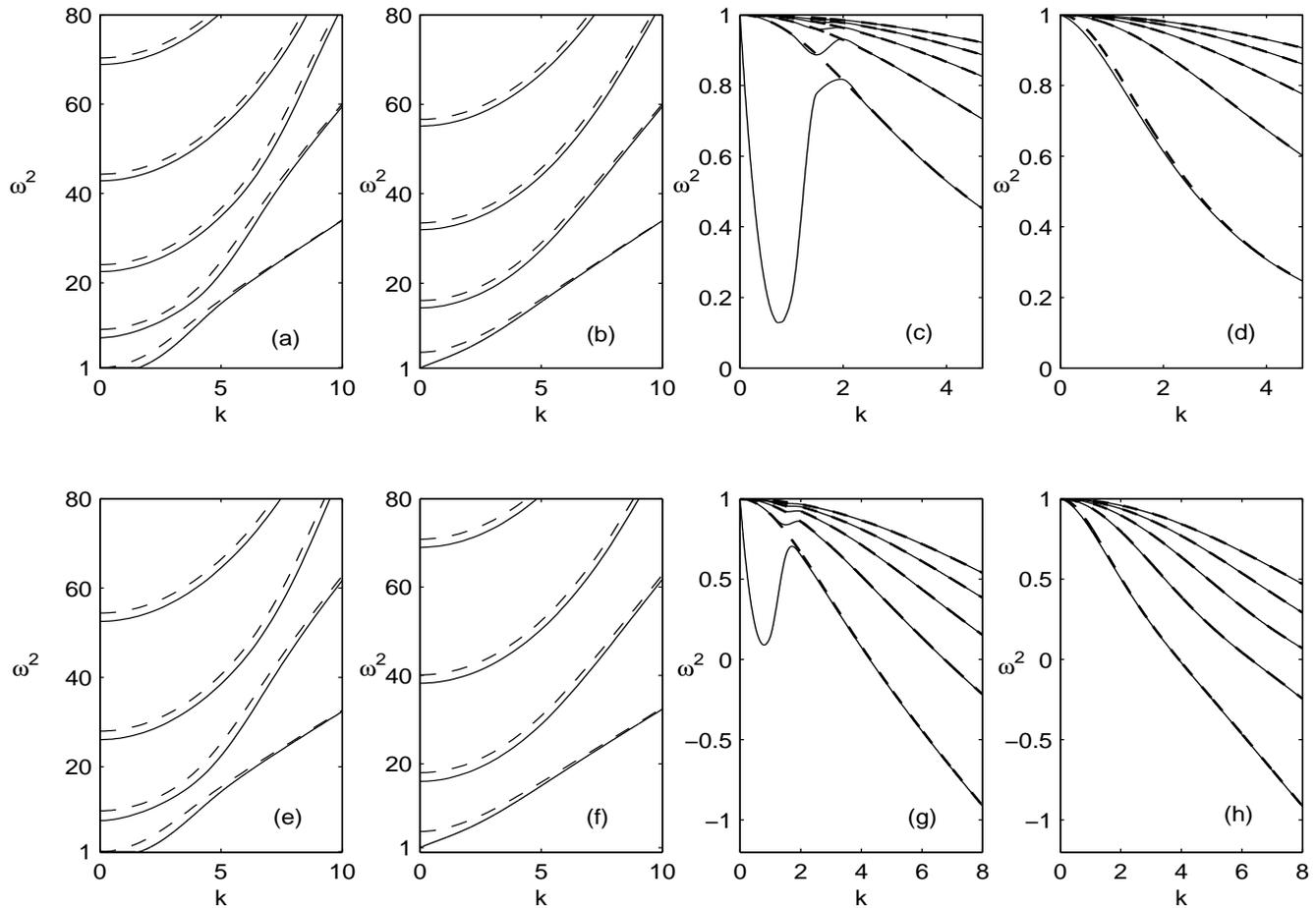}
\caption{Dispersion diagrams for vertical modes in a
self-gravitating disc with $Q=0.2$ for an adiabatic stratification
with $s=2.5$ (panels a-d) and for a superadiabatic stratification
with $s=1.7$ (panels e-h). Shown are (a,e) the even p- and f-modes,
(b,f) the odd p- and f-modes, (c) the even r-mode, (d) the odd
r-mode, (g) the convectively unstable even g-mode and (h) the odd
g-mode. The frequency ordering and mode naming are the same as in
Fig. 2. As before, dashed curves correspond to the dispersion
diagrams calculated without self-gravity in the perturbations. The
behaviour of the p-, f- and r-modes in the adiabatic case is similar
to that in Fig. 3. In the superadiabatic case, the dispersion curves
of the even g-mode have dips caused by self-gravity in the interval
$0 \leq k \leq 2$, while the odd g-mode is almost unaffected by
self-gravity. The largest dip occurs for the basic branch of the
even g-mode (denoted above as ${\rm g}_{0}^{e}$) in panel (g), which
appears to be most significantly influenced by self-gravity. For
large $k$ and/or large mode numbers, the effect of self-gravity
becomes weak and the dispersion curves merge with dashed ones for
the non-self-gravitating case. Due to the superadiabatic
stratification, the even and odd g-modes are convectively unstable
(i.e., have $\omega^2<0$) in the range $k>3.91$, where the influence
of self-gravity is small.}
\end{figure*}

\subsection{Superadiabatic stratification}

The superadiabatic case (Figs. 5e-5h) is interesting, because at
$\omega^2\leq \kappa^2$, as classified above, instead of the r-mode
there is the g-mode, which in addition to convective instability can
also exhibit gravitational instability. As in the
non-self-gravitating case, for radial wavenumbers larger than a
certain value, all the branches of the g-mode become unstable
because of the negative vertical entropy gradient. From Fig. 5g we
see that for $0\leq k \leq 2$, self-gravity produces dips on the
g-mode dispersion curves, again preferably for the even parity ones.
When $Q$ drops sufficiently, the dip on the basic branch of the even
g-mode, which appears to be most affected by self-gravity, starts to
cross the $\omega^2=0$-axis in a way similar to that of the basic
even r-mode branch above and this signals the onset of gravitational
instability. For $s=1.7$, adopted in Fig. 5, we find $Q_{cr}=0.184$,
which gives a thickness $H=2h=1.76$, and $k_{m}=0.83$, so that
$\lambda_{m}=2\pi/k_{m}= 7.57 \gsim H$. As for the p- and f-modes,
they behave under self-gravity in exactly the same manner as in the
above cases. In particular, although their frequencies are reduced,
they can never become gravitationally unstable even for very small
values of $Q$. Thus, for superadiabatic stratification, the basic
even g-mode can exhibit two types of instabilities: gravitational
and convective (see Fig. 6). As we see from Figs. 5g,6a, at moderate
$Q\sim Q_{cr}$, the radial scales for the activity of self-gravity
and convective instability are well separated, so that these two
effects do not interfere with each other in the linear regime. For
$Q=0.1$ in Fig. 6a, self-gravity is dominant at $0 \leq k \leq
2.55$, while convective instability occurs at $k>4.27$, where the
effect of self-gravity is weak (a similar situation is for $Q=0.2$
in Fig. 5g). However, for very small $Q\ll Q_{cr}$, the radial
scales of gravitational and convective instabilities overlap (Fig.
6b), but the gas motion for such radial scales in the case of very
strong gravitational instability would hardly resemble convective
motions (it will look more like that in Fig. 11 below); convection
is simply disrupted by gravitational instability. Therefore, we can
conclude that unless a disc is strongly self-gravitating,
self-gravity has little influence on the properties of convective
motions and on the (Schwarzschild) criterion for the onset of
convective instability that is equivalent to $N_0^2<0$.

\begin{figure}
\centering\includegraphics[width=\columnwidth]{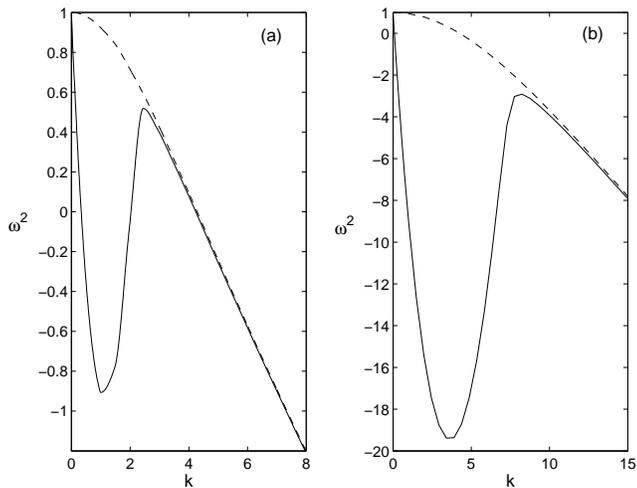}
\caption{Dispersion curves of the basic even g-mode for a
superadiabatic stratification with $s=1.7$ for (a) $Q=0.1$ and (b)
$Q=0.01$. Dashed lines show the same branches computed without
self-gravity in the perturbations. For $Q=0.1$, the basic even
g-mode exhibits gravitational instability in the vicinity of
$k_{m}=1.15$ and convective instability in the range $k>4.27$, where
the influence of self-gravity is weak. For $Q=0.01$, the effect of
self-gravity is much stronger and the radial scales of gravitational
and convective instabilities overlap in the range $4.25 \leq k \leq
8.5$.}
\end{figure}

\subsection{Analogy with the 2D density wave theory}

From the three cases considered above, we conclude that the basic
branch of the low-frequency ($\omega^2\leq \kappa^2$) even mode,
which is the r-mode, in the case of subadiabatic and adiabatic
stratifications, and the g-mode, in the case of superadiabatic
stratification, is most subject to the influence of self-gravity.
These modes determine the gravitational instability in vertically
stratified discs. As we have clearly seen in all the above cases,
the radial scale of the instability is comparable, though a bit
larger than the disc thickness, $\lambda_{m}\gsim H$, which in turn
implies that a 2D analysis of the gravitational instability is at
most marginally valid and the rigorous stability (linear) analysis
of self-gravitating discs should be three-dimensional. However,
despite the 2D treatment only being marginal, we can still find
similarities between our results and the 2D density wave theory in
thin self-gravitating discs, which actually appears to do a decent
qualitative job.

Consider an equivalent zero-thickness disc with the sound speed
$c_{sm}$\footnote{In fact, there are other options for choosing the
sound speed as a some kind of height-averaged value. We address this
question in Section 6, but for the present purpose this does not
make much difference.} and the surface density
$\Sigma_0=\int_{-h}^h\rho_0 dz$. Then the well-known dispersion
relation of axisymmetric 2D density waves in the thin disc is
\cite[e.g.,][]{GT78}
\[
\omega^2=k^2-\frac{2}{Q_{2D}}k+1,
\]
where we have used the same normalization as before and because of
that $Q_{2D}=c_{sm}\Omega/\pi G \Sigma_0$. This dispersion relation
is a parabola with a minimum at the Jeans wavenumber $k_J=1/Q_{2D}$.
This is the wavenumber, at which the effect of self-gravity is most
prominent, and if $Q_{2D}<1$ it gives the characteristic scale of
gravitational instability. At small $k\ll k_J$, 2D density waves are
dominated by self-gravity and inertial forces, while at large $k\gg
k_J$, pressure/compressibility dominates over self-gravity and
density waves appear as an acoustic mode. At $k\sim k_J$ all three
factors can be important. Let us now look at the dispersion curves
of the basic branches of the even r- and convectively unstable even
g-modes in Figs. 3c,5c,5g. They have similar parabolic shape in the
self-gravity and compressibility dominated regime at $k \sim k_m$
with the linear phase at smaller $k\ll k_m$, where only self-gravity
and inertial forces play a role. This linear phase at long
wavelengths is well reproduced by the 2D dispersion relation.
Therefore, we can identify the wavenumber $k_m$, at which the effect
of self-gravity on the 3D modes is largest, with the Jeans
wavenumber $k_J$. In some sense, in the case of instability when
$Q<Q_{cr}$, $k_m$ gives a more accurate value for the radial scale,
$\lambda_m$, of the gravitationally most unstable mode than that
given by the Jeans wavenumber $k_J$ in the 2D model. For example,
from Fig. 12 we find that the vertical structure ($\rho_0$)
calculated at $s=5$ and $Q_{cr}=0.168$ gives the corresponding
$Q_{2D}=0.76$ and $k_J=1.32$, which differs from $k_m=0.8$ found
above for these parameters. Moreover, we will see in Section 6 that
the criterion for gravitational instability based on the 3D
calculations, i.e. $Q<Q_{cr}$, is more exact than the 2D criterion
$Q_{2D}<1$.

\begin{figure}
\centering\includegraphics[width=\columnwidth]{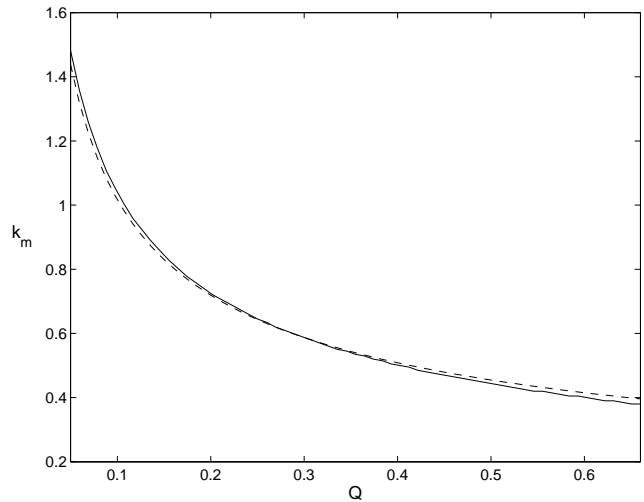}
\caption{Dependence of $k_m$ on $Q$ for $s=5$, which very closely
follows the power law $Q^{-1/2}$ (shown with dashed line and scaled
appropriately). The gravitational instability sets in at
$Q=Q_{cr}=0.168$ that gives $k_m=0.8$.}
\end{figure}

In the 2D case, as shown above, the Jeans wavenumber $k_J=1/Q_{2D}$,
implying that it is determined solely by competition between
self-gravity and compressibility/pressure. It is now interesting to
see how an analogous characteristic wavenumber $k_m$ depends on $Q$
in the 3D case. From Fig. 7 we see that this dependence has a power
law character $Q^{-1/2}$, which means that the value of $k_m$ is
again set by self-gravity and compressibility, as in the 2D case, so
that the disc rotation plays a role only in determining $\omega^2$
but not $k_m$. Indeed, the only lengthscale that can be constructed
from the sound speed $c_{sm}$, density $\rho_m$ and gravitational
constant $G$ without the angular velocity, $\Omega$, of disc
rotation is $c_{sm}/\sqrt{G\rho_m}$ (analogous to the Jeans length
of a collapsing 3D cloud). If we normalize it by $c_{sm}/\Omega$, we
get $2\sqrt{\pi}Q^{1/2}$, giving the above power law dependence for
the corresponding non-dimensional wavenumber $\sqrt{\pi}Q^{-1/2}$ to
which $k_m$ is proportional.

Thus, the basic even r-mode or, in the case of superadiabatic
stratification, the basic even g-mode at $k \lsim k_m$ exhibit many
of the characteristics of 2D density waves and could be regarded as
their 3D analogues at such radial wavenumbers (see also Section 5).
If continued to $k\gg k_m$, the density wave mode would thus connect
up to the large-$k$ parts of these two modes dominated,
respectively, by inertial forces and by negative buoyancy, instead
of merging with the compressible p-mode, as might seem at first
sight from the 2D dispersion relation. However, the limit $k\gg
k_m$, equivalent to $\lambda \ll H$, means the breakdown of the 2D
approximation and, therefore, the concept of 2D density waves at
$k\gg k_m$ would not be well-defined.

\section{Gravitational instability in 3D: properties of the basic
branch of the low-frequency even r-mode}

\begin{figure}
\centering\includegraphics[width=\columnwidth]{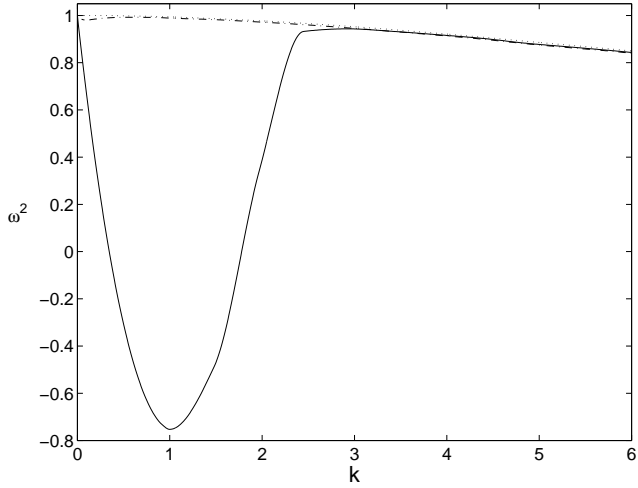}
\caption{Dispersion curves of the basic even r-mode at $s=5,Q=0.1$
with corresponding $h=1.22$. Dotted line corresponds to the
non-self-gravitating case, dashed-dotted -- to the incompressible
case with self-gravity and solid line -- to the self-gravitating
case with compressibility. The influence of self-gravity on the
basic even r-mode is very weak in the incompressible limit and the
dashed-dotted dispersion curve almost coincides with the dotted one
in the non-self-gravitating limit; a slight deviation (dip) at small
$k$ is due to the stratified background.}
\end{figure}

In this section we concentrate only on the properties of the
gravitationally unstable basic even r-mode. As we have seen, the
behaviour of the convectively unstable basic even g-mode under
self-gravity has a qualitatively similar character.

\subsection{Effect of compressibility}
As noted above, in the Poisson's equation, the density perturbations
giving rise to the gravitational potential perturbations can be due
to compressibility and due to the stratification of the equilibrium
vertical structure. As is evident from Fig. 2c, in
non-self-gravitating discs, the r-mode is nearly incompressible. So,
it is interesting to see if it still remains incompressible in
self-gravitating discs and which out of these two sources of density
perturbations is ultimately responsible for the gravitational
instability. In order to explore this, we again computed the
dispersion curve of the basic even r-mode in the incompressible
limit of equations (16-18), which take the form:
\begin{equation}
\frac{du_z}{dz}=\frac{k^2}{\omega^2-\kappa^2}(p+\psi)
\end{equation}
\begin{equation}
\frac{dp}{dz}=\frac{N_0^2}{g_0}p+(\omega^2-N_0^2)u_z-\frac{d\psi}{dz}
\end{equation}
\begin{equation}
\frac{d^2\psi}{dz^2}-k^2\psi=\frac{\rho_0}{Q}\frac{N_0^2}{g_0}u_z.
\end{equation}
supplemented with the same boundary conditions. Notice that on the
right hand side of Poisson's equation (25), only the density
perturbation due to stratification remains. In Fig. 8, we compare
the dispersion curves of the basic even r-mode obtained in the
incompressible limit to those computed for the compressible
self-gravitating and non-self-gravitating cases. It is clear that in
self-gravitating discs, the basic even r-mode becomes compressible
at $k \sim k_m$ thereby providing density perturbation for the
gravitational potential and, therefore, accounting for the large dip
on the dispersion curve. The dispersion curve in the incompressible
limit deviates only very slightly from that in the
non-self-gravitating case, because of the second source of density
perturbation, i.e. stratification, which turns out to be much
smaller than that associated with compressibility. Such a comparison
also shows that the primary cause of gravitational instability in
incompressible discs, as described in GLB and \cite{LP93b}, is the
displacements of free-surfaces. In our case, the equilibrium density
vanishes at the surface, so that the effect of surface displacements
on the growth rate of instability is null (see boundary condition
22) and in the incompressible case only density perturbation due to
stratification, as shown, is hardly sufficient for gravitational
instability.

\begin{figure}
\includegraphics[width=\columnwidth]{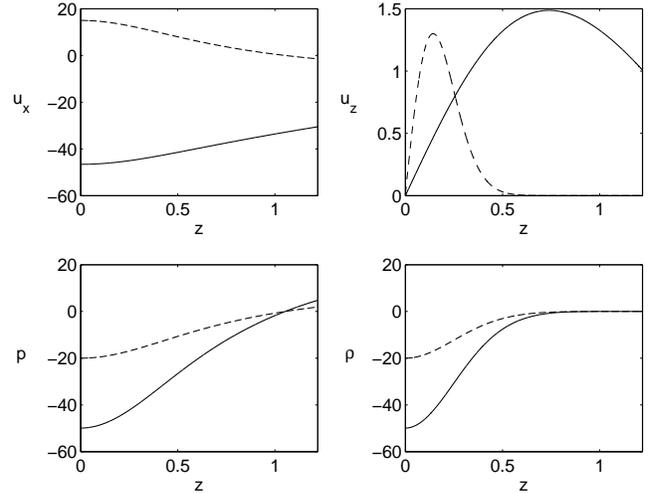}
\caption{Vertical structures of the radial velocity $u_x$, the
vertical velocity $u_z$, the pressure $p$ and the density $\rho$
perturbations constituting the eigenfunctions of the gravitationally
unstable basic even r-mode for $s=5, Q=0.1, h=1.22, k_{m}=1$ with
the corresponding largest growth rate $\omega_{min}^2=-0.75$. Dashed
lines show the eigenfunctions (scaled arbitrarily for plotting
purposes) in the non-self-gravitating case for the same parameters
except $\omega^2=0.99$. Notice that in the non-self-gravitating
case, the eigenfunctions are more concentrated near the midplane,
whereas in the self-gravitating case they vary over the whole
vertical extent.}
\end{figure}

\begin{figure}
\includegraphics[width=\columnwidth]{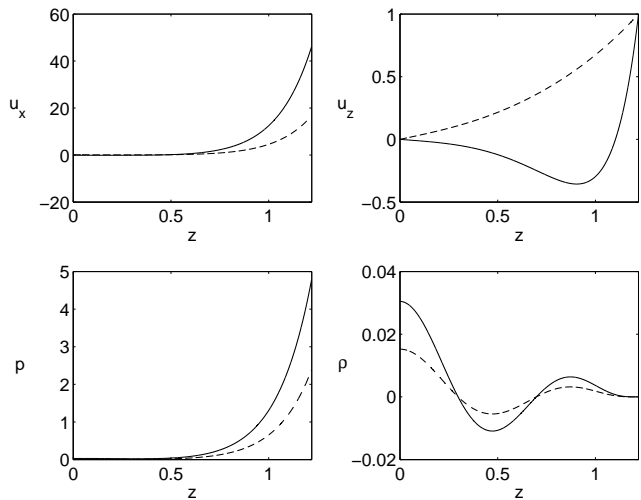}
\caption{Eigenfunctions of the even f-mode, which unlike the basic
even r-mode is gravitationally stable, with the same values of
parameters as in Fig. 9. The corresponding eigenfrequencies at
$k_{m}=1$ are $\omega^2=1.1$ for the self-gravitating case (solid
lines) and $\omega^2=1.6$ for the non-self-gravitating case (dashed
lines). Again, non-self-gravitating eigenfunctions have been scaled
arbitrarily for plotting purposes.}
\end{figure}

\subsection{Eigenfunctions and spatial structure}
Here we compute the vertical structure of the eigenfunctions of the
gravitationally unstable basic even r-mode and also find what type
of motions it induces. We take $Q=0.1$ and $s=5$, which gives
$h=1.22$ for the disc height. For these parameters, the dispersion
curve of this mode in Fig. 8 has a minimum $\omega_{min}^2=-0.75$,
i.e. gives the largest growth rate of gravitational instability, at
$k_{m}=1$. In Fig. 9, we plot the corresponding eigenfunctions only
in the upper half of disc's full vertical extent. Because the mode
has even parity, the vertical velocity and the derivative of the
gravitational potential are odd functions and, correspondingly, the
pressure and potential are even functions of $z$. Only the pressure
perturbation has one node in the interval $0 < z \leq h$, other
quantities have no nodes and vary over the whole vertical extent.
This demonstrates that the spatial structure of the gravitationally
most unstable mode is three-dimensional with non-zero vertical
velocity. To see how self-gravity changes the form of the
eigenfunctions, we also show the eigenfunctions of the same branch
in the non-self-gravitating limit, which appear to be more
concentrated towards the midplane than the self-gravitating ones and
have no nodes in the same vertical range. Since the f-mode plays an
important role in the angular momentum and energy transport in
stratified discs (LO98), in Fig. 10 we also plot its eigenfunctions
in the presence of self-gravity for the same parameters. Comparing
Figs. 9 and 10, we clearly see that self-gravity modifies the form
of the eigenfunctions of both the f- and r-modes. The perturbed
quantities for both mode types vary with height somewhat similarly,
especially the vertical velocities, which are, however, still at
least an order of magnitude less than the horizontal ones, so that
the motions can be viewed, to leading order, as 2D (see also Fig.
11). Thus, the gas motion associated with the gravitationally
unstable r-mode is of similar type to that of the f-mode, which, in
turn, implies that the r-mode might be as important as the f-mode,
i.e., might be able to do similar `dynamical jobs' as the f-mode in
self-gravitating discs. As mentioned earlier, the non-linear
behaviour of the 3D perturbations in self-gravitating discs has been
attributed solely to the f-mode dynamics without analysing other
modes \citep{Petal00,BD06}.

The perturbed quantities have the form $f(z)exp(ik_{m}x), f\equiv
(u_x,u_z,p,\rho)$ and from this we can construct the spatial picture
of the velocity, density and temperature perturbations for the
gravitationally unstable basic even r-mode by taking the real parts.
The temperature perturbation is found from the ideal gas equation of
state and is equal to $T=-c_s^2\rho/\rho_0+\gamma p/\rho_0$ (here
the pressure perturbation $p$ is as used in the original equations
10-15, i.e., before switching to new variables). Figure 11 shows the
density and temperature fields constructed in this way with velocity
vectors showing the direction of gas motion superimposed on the
density field, which in fact resembles the classical density profile
of a 2D density wave. Near the midplane, matter flows (converges),
almost parallel to the $x-$axis, towards high density regions. With
increasing $z$, the flow becomes more arc-like, because of the
variation of the vertical velocity with height. An analogous
velocity field in the $(x,z)$ plane was also observed in some
non-linear simulations of self-gravitating discs
\citep{BD06,Betal06}, which suggests that the latter may be a result
of the non-linear development of the type of motion we see here in
the linear regime. We might also speculate that in relatively
high-mass discs, the non-axisymmetric gravitationally unstable basic
r-mode will produce similar high-density regions, which can be
precursors of spiral shock fronts (if the disc does not fragment),
and arc-like motions in the upper layers that can give rise to
non-linear shock bores involving large distortions of the disc
surface as described by \cite{BD06}. The temperature perturbations
are fairly non-uniform as well, varying both vertically and radially
on comparable scales. Obviously, larger temperature perturbations
correspond to overdense regions that are contracting because matter
flows into them, and lower -- to underdense regions.

\begin{figure}
\centering\includegraphics[width=\columnwidth]{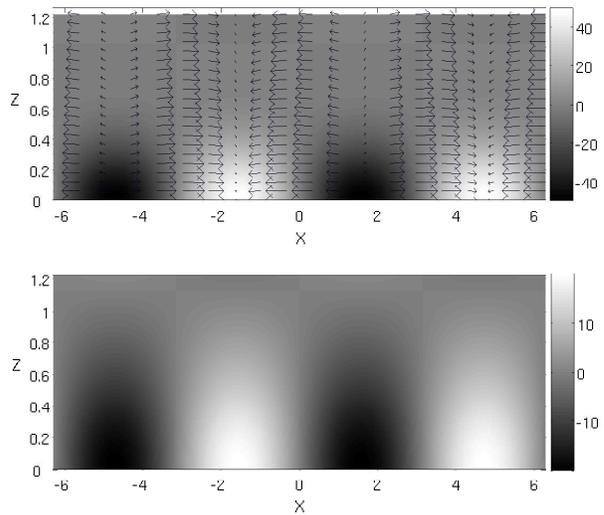}
\caption{Distribution of the density (upper plot) and temperature
(lower plot) perturbations in $(x,z)$ plane constructed from the
eigenfunctions of the gravitationally unstable basic even r-mode in
Fig. 9. Superimposed on the density field are the velocity vectors
of the induced gas flow.}
\end{figure}

\section{Stability criteria in 2D and 3D}

\begin{figure}
\centering\includegraphics[width=\columnwidth]{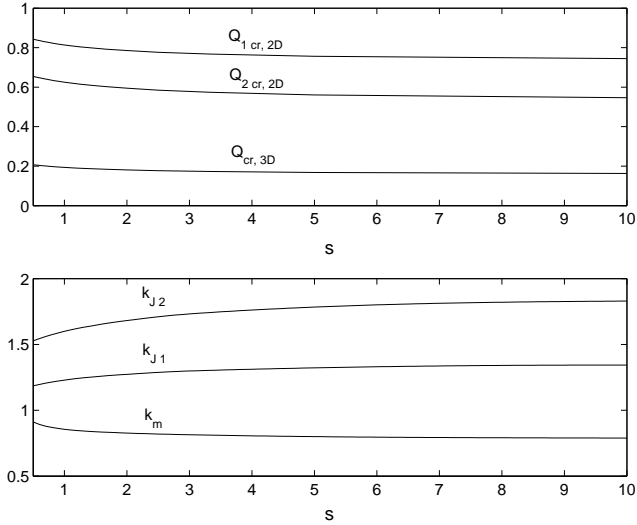}
\caption{Dependence of the critical $Q_{cr,3D}$ and the
corresponding critical $Q_{1~cr,2D}$ and $Q_{2~cr,2D}$ on the
polytropic index $s$. Shown also are the $k_m$ at $Q_{cr,3D}$ and
the 2D Jeans wavenumbers
$k_{J1}=1/Q_{1~cr,2D},~k_{J2}=1/Q_{2~cr,2D}$ as a function of $s$.}
\end{figure}

In the 3D case, the degree and effect of self-gravity is
characterized by $Q_{3D}=\Omega^2/4\pi G \rho_m$, where $\rho_m$ is
the value of equilibrium density at the midplane (see also GLB),
which plays here a role similar to that of the standard 2D Toomre's
parameter $Q_{2D}=c_s\Omega/\pi G \Sigma$ \citep{T64}. However, in
three-dimensional simulations, researchers often still tend to use
the 2D Toomre's parameter to characterize the onset of gravitational
instability \citep{Riceetal03,RLA05,LR04,LR05,Betal06}. In this
section we investigate how a more general and exact 3D criterion of
gravitational instability can be related to the 2D one. In general,
these two parameters are different, because the 2D parameter
contains the sound speed and surface density, which by definition do
not vary vertically for razor-thin discs, whereas the 3D parameter
does not contain the sound speed and surface density, but the
midplane value of the density. So, if we still want to characterize
the 3D instability in terms of the 2D Toomre's parameter, we need to
use some height-averaged, or effective sound speed (surface density
can be uniquely calculated from the vertical density distribution).
In some simulations the midplane values \citep{Metal05,Betal06},
while in others vertically averaged values \citep{Riceetal03,RLA05}
of the sound speed are used. Here we calculate the corresponding
critical $Q_{cr, 2D}$ for these cases.

As is well-known, in razor-thin 2D discs, axisymmetric perturbations
(density waves) are gravitationally unstable if $Q_{2D}<1$
\citep{T64}. Let us now find the critical $Q_{cr,3D}$ that
determines the gravitational instability in 3D, which, as found in
this study, is associated with the basic branches of the even parity
low-frequency modes. In Fig. 12, we show the critical $Q_{cr,3D}$ as
a function of the polytropic index $s$. Given $Q_{cr,3D}$, we can
now find $Q_{1~cr,2D}$ defined in terms of the midplane sound speed
and $Q_{2~cr,2D}$ defined in terms of the vertically averaged sound
speed:
$$
Q_{1~cr,2D}=\frac{2Q_{cr,3D}}{\int_0^h \rho_0 dz},\\~~~~\\
Q_{2~cr,2D}=\frac{2Q_{cr,3D}\int_0^h c_s dz}{h \int_0^h \rho_0 dz},
$$
where $\rho_0$ and $c_s$, as before, are the normalized equilibrium
density and sound speed the $z$-dependence of which, as well as the
value of $h$, are determined by $Q_{cr,3D}$ and the polytropic index
$s$. So, the critical $Q_{1~cr,2D}$ and $Q_{2~cr,2D}$ calculated
with the two different methods are different, though not far from
each other, and less than unity which is their critical value in the
2D case. This again confirms the well-known result that 3D discs are
more stable and to make them unstable smaller value of $Q_{2D}$ is
necessary when compared with that required for razor-thin 2D discs
(see also GLB). But in such cases, using $Q_{3D}$ seems more
appropriate as it does not involve uncertainties in how to average
the sound speed over height.

In Fig. 12 we also examine the dependence of the critical
wavenumbers of the gravitational instability on $s$. We see that the
2D Jeans wavenumbers are always larger than the actual more exact
wavenumber $k_m$ at the onset of gravitational instability when
$Q=Q_{cr,3D}$. Thus, in the 3D case, the critical wavenumbers depend
differently on the thermodynamics than in the 2D case, where the
critical $k_J=1/Q_{2D}=1$ at the onset of instability.

\section{Summary and discussions}

In this paper, we have analysed the axisymmetric normal modes of
perturbations in stratified, compressible, self-gravitating gaseous
discs with subadiabatic, adiabatic and superadiabatic vertical
stratifications. First, we performed a classification of
perturbation modes in stratified discs in the absence of
self-gravity to compare with previous calculations. Four well-known
main types of modes can be distinguished: acoustic p-modes, surface
gravity f-modes, buoyancy g-modes and inertial r-modes. The
restoring forces for these modes for large radial wavenumbers are
mainly provided by one the following: pressure/compressibility,
displacements of the disc surface, buoyancy and inertial forces due
to disc rotation for the p-,f-,g- and r-modes, respectively. For
smaller wavenumbers, the restoring force for each mode is a
combination of these forces. In the case of adiabatic
stratification, buoyancy is zero and, therefore, the g-mode
disappears, while other modes remain qualitatively unchanged. For
superadiabatic stratification, the g-mode becomes convectively
unstable and merges with the r-mode, so that only a single
convectively unstable mode appears in the dispersion diagram at
$\omega^2 \leq \kappa^2$, which we still call the g-mode. Due to the
reflection symmetry of the equilibrium vertical structure with
respect to the midplane, each mode comes in even and odd pairs. By
our terminology, for even (odd) modes, pressure and gravitational
potential perturbations are even (odd), while the perturbations of
vertical velocity and derivative of potential are odd (even)
functions of the vertical coordinate. After classifying and
characterizing modes in the absence of self-gravity, we introduced
self-gravity in the perturbation equations and investigated how it
modifies the properties of these modes. We found that self-gravity,
to some extent, reduces the frequencies of all normal modes at
radial wavelengths comparable to the disc height, but its influence
on the basic even r-mode, in the case of subadiabatic and adiabatic
stratifications, and on the basic even g-mode, in the case of
superadiabatic stratification, appears to be strongest. With
decreasing $Q_{3D}$, these modes become unstable due to self-gravity
and thus determine the gravitational instability of a vertically
stratified disc. The basic even g-mode also exhibits convective
instability due to a negative entropy gradient but, unless the disc
is strongly self-gravitating, these two instabilities grow
concurrently in the linear regime, because their corresponding
radial scales are separated. We also obtained the corresponding
criterion for the onset of gravitational instability in 3D, which is
more exact than the standard instability criterion in terms of the
2D Toomre's parameter, $Q_{2D}<1$, for axisymmetric density waves in
razor-thin discs. By contrast, the p-, f- and convectively stable
g-modes have their $\omega^2$ reduced by self-gravity, but never
become unstable for any value of $Q_{3D}$. This is a consequence of
the three-dimensionality of the disc. The eigenfunctions associated
with the gravitationally unstable modes are intrinsically
three-dimensional, that is, have non-zero vertical velocity and all
perturbed quantities vary over the whole vertical extent of the
disc. In this regard, we would like to mention that resolving the
gravitationally most unstable mode in numerical simulations thus
reduces to properly resolving the disc height (together with
resolving the corresponding radial scale, which, as shown here,
appears somewhat larger than the disc height). So, the criterion of
\citet{N06} that at least $\sim 4$ particle smoothing lengths should
fit into per scale height may apply in three-dimensional SPH
simulations. This implies a substantially larger number of SPH
particles per vertical column because the disc itself may extend
over many scale heights. He also shows that a similar criterion
applies to grid-based simulations.

Here for simplicity, and also to gain the first insight into the
effects of self-gravity on the vertical modes in stratified discs,
we considered only axisymmetric perturbations. Non-axisymmetric
perturbations are dynamically richer, though more complicated,
because the phenomena induced by Keplerian shear/differential
rotation -- strong transient amplification of perturbations and
\emph{linear} coupling of modes (not to be confused with non-linear
mode-mode interactions) -- come into play for these type of
perturbations. Transient (swing) amplification of perturbations
(density waves) has been studied previously in razor-thin 2D
approximation \citep{GLB65b,GT78,T81,MC07}. From the analysis
presented here, we may expect that in the linear regime, the
non-axisymmetric basic even r-mode can undergo larger transient
amplification due to self-gravity than other modes in the disc. This
transient amplification of perturbations may be important for
explaining the large burst phases seen in numerical simulations at
the initial stages of the development of gravitational instability
in discs \citep{Riceetal03,RLA05,LR04,LR05,Betal06}. So, in this
respect one should analyse and quantify the transient amplification
of non-axisymmetric perturbations in stratified 3D self-gravitating
discs starting with linear theory. As for the linear coupling of
modes, it was demonstrated that in non-self-gravitating stratified
discs, Keplerian shear causes rotational (vortex) mode perturbations
to couple with and generate g-mode perturbations \citep{TCZ08}. In
the context of 2D discs, it was shown that vortex mode perturbations
can also excite density waves due to shear
\citep{Bodoetal05,MC07,MR09,HP09}. In the 3D case, there are a
larger number of modes in a disc and it is quite possible that some
of them may appear linearly coupled due to shear and, therefore, be
able to generate each other during evolution, especially the f-mode
and the basic branch of the r-mode (because they vary on comparable
vertical scales in the presence of self-gravity, Figs. 9, 10).
Another related problem also of interest is the interaction between
self-gravity and the MRI in magnetized discs \cite[see
e.g.,][]{F05}. In particular, how self-gravity can modify the growth
rates of magnetic normal modes responsible for the MRI. Actually,
this will be the generalization of the extensive analysis of normal
modes in magnetized discs by \cite{Og98}.

\cite{LO98} showed that in non-self-gravitating discs with
polytropic vertical stratification, an external forcing
preferentially excites the f-mode, because it has the largest
responsiveness to an external driving compared to other modes. This
mode, propagating through a disc, results in energy dissipation near
the disc surface. Based on these results and partly on the
properties of f-modes in stellar dynamics, \citet{Petal00}
identified the behaviour of 3D perturbations in self-gravitating
discs involving large surface distortions and the resulting energy
dissipation in the upper layers, with f-mode dynamics. The
self-stimulated potential was thought to play the role of an
external/tidal force. However, it is not obvious that the effect of
a self-stimulated potential is the same as that of the external
potential. In fact, our analysis has revealed that in
self-gravitating discs, in addition to the f-mode, the r-mode can
also be dynamically important, because this mode appears to be
subject to gravitational instability, while the f-mode is not. The
eigenfunctions of the gravitationally unstable basic even r-mode
differ from those of the r-mode in non-self-gravitating discs in
that they are no longer concentrated near the midplane and behave
somewhat similarly to the eigenfunctions of the f-mode: they vary
over the whole vertical extent of the disc and also involve
noticeable perturbations of the disc surface. Consequently, like the
f-mode, the gravitationally unstable r-mode can, in principle, also
induce gas motion causing large surface distortions and resultant
energy dissipation in the upper layers of the disc, which is thought
to play a role in enhancing disc cooling \cite[because the energy is
deposited in the upper layers with small optical depth, it can be
radiated away more quickly and effectively cool the disc, but this
is a subject of further study, see e.g.,][]{JG03,Betal06}.
Furthermore, in the case of non-axisymmetric perturbations, as
mentioned above, because of shear, the gravitationally unstable
r-mode can couple with and generate the strong f-mode. So, the
surface distortions may be caused by a combination of the f- and
r-modes. In order to explore where dissipation can predominantly
occur in a self-gravitating disc, one needs to generalize the
analysis of \cite{LPS90}, LP, LO98, \cite{Bateetal02} on the
propagation of waves in stratified non-self-gravitating discs and
consider the propagation of non-axisymmetric modes in stratified
self-gravitating discs. The dispersion properties of modes in the
presence of self-gravity as found here are one of the necessary
things for studying mode propagation.

Another point we want to raise concerns the spatial distribution of
temperature. In order to realistically model the cooling of
protoplanetary discs, \citet{Betal07} employed the radiative
transfer technique. In the vertical $z-$direction, the radiative
transfer equation was solved exactly assuming a plane-parallel
atmosphere approximation, but in the radial direction only the
radiation diffusion approximation was employed. However, as our
linear results (Fig. 11) and other non-linear simulations
\citep[e.g.,][]{Metal05,Betal06} demonstrate, temperature and,
therefore, opacity may vary on comparable scales in both the radial
and vertical directions and have very complex structure in the
non-linear regime. This implies that a more general radiative
transfer treatment based on solving the ray equation in all
directions, rather than using the diffusion approximation in either
direction, would be more appropriate for better understanding
cooling processes.

\section*{Acknowledgments}
G.R.M. would like to acknowledge the financial support from the
Scottish Universities Physics Alliance (SUPA). We thank Gordon
Ogilvie for helpful discussions and for critically reading the
manuscript. We also thank the referee for the constructive comments
that improved the presentation of our results.

\bibliographystyle{mn2e}
\bibliography{bibliography}

\begin{thebibliography}{}

\bibitem[\protect\citeauthoryear{{Adams}, {Ruden} \& {Shu}}{{Adams}
  et~al.}{1989}]{ARS89}
{Adams} F.~C.,  {Ruden} S.~P.,    {Shu} F.~H.,  1989, ApJ, 347, 959

\bibitem[\protect\citeauthoryear{{Bate}, {Ogilvie}, {Lubow} \&
  {Pringle}}{{Bate} et~al.}{2002}]{Bateetal02}
{Bate} M.~R.,  {Ogilvie} G.~I.,  {Lubow} S.~H.,    {Pringle} J.~E.,  2002,
  MNRAS, 332, 575

\bibitem[\protect\citeauthoryear{{Bertin}, {Lin}, {Lowe} \&
  {Thurstans}}{{Bertin} et~al.}{1989}]{Bertetal89}
{Bertin} G.,  {Lin} C.~C.,  {Lowe} S.~A.,    {Thurstans} R.~P.,  1989, ApJ,
  338, 104

\bibitem[\protect\citeauthoryear{{Binney} \& {Tremaine}}{{Binney} \&
  {Tremaine}}{1987}]{BT87}
{Binney} J.,  {Tremaine} S.,  1987, {Galactic dynamics}.
Princeton University Press, Princeton, NJ

\bibitem[\protect\citeauthoryear{{Bodo}, {Chagelishvili}, {Murante},
  {Tevzadze}, {Rossi} \& {Ferrari}}{{Bodo} et~al.}{2005}]{Bodoetal05}
{Bodo} G.,  {Chagelishvili} G.,  {Murante} G.,  {Tevzadze} A.,  {Rossi} P.,
  {Ferrari} A.,  2005, A\&A, 437, 9

\bibitem[\protect\citeauthoryear{{Boley}}{{Boley}}{2009}]{Bol09}
{Boley} A.~C.,  2009, ApJ, 695, L53

\bibitem[\protect\citeauthoryear{{Boley} \& {Durisen}}{{Boley} \&
  {Durisen}}{2006}]{BD06}
{Boley} A.~C.,  {Durisen} R.~H.,  2006, ApJ, 641, 534

\bibitem[\protect\citeauthoryear{{Boley}, {Durisen}, {Nordlund} \&
  {Lord}}{{Boley} et~al.}{2007}]{Betal07}
{Boley} A.~C.,  {Durisen} R.~H.,  {Nordlund} {\AA}.,    {Lord} J.,  2007, ApJ,
  665, 1254

\bibitem[\protect\citeauthoryear{{Boley}, {Mej{\'{\i}}a}, {Durisen}, {Cai},
  {Pickett} \& {D'Alessio}}{{Boley} et~al.}{2006}]{Betal06}
{Boley} A.~C.,  {Mej{\'{\i}}a} A.~C.,  {Durisen} R.~H.,  {Cai} K.,  {Pickett}
  M.~K.,    {D'Alessio} P.,  2006, ApJ, 651, 517

\bibitem[\protect\citeauthoryear{{Boss}}{{Boss}}{1998}]{B98}
{Boss} A.~P.,  1998, ApJ, 503, 923

\bibitem[\protect\citeauthoryear{{Boss}}{{Boss}}{2003}]{B03}
{Boss} A.~P.,  2003, ApJ, 599, 577

\bibitem[\protect\citeauthoryear{{Boss}}{{Boss}}{2004}]{B04}
{Boss} A.~P.,  2004, ApJ, 610, 456

\bibitem[\protect\citeauthoryear{{Fromang}}{{Fromang}}{2005}]{F05}
{Fromang} S.,  2005, A\&A, 441, 1

\bibitem[\protect\citeauthoryear{{Gammie}}{{Gammie}}{2001}]{Gam01}
{Gammie} C.~F.,  2001, ApJ, 553, 174

\bibitem[\protect\citeauthoryear{{Goldreich} \& {Lynden-Bell}}{{Goldreich} \&
  {Lynden-Bell}}{1965a}]{GLB65a}
{Goldreich} P.,  {Lynden-Bell} D.,  1965a, MNRAS, 130, 97

\bibitem[\protect\citeauthoryear{{Goldreich} \& {Lynden-Bell}}{{Goldreich} \&
  {Lynden-Bell}}{1965b}]{GLB65b}
{Goldreich} P.,  {Lynden-Bell} D.,  1965b, MNRAS, 130, 125

\bibitem[\protect\citeauthoryear{{Goldreich} \& {Tremaine}}{{Goldreich} \&
  {Tremaine}}{1978}]{GT78}
{Goldreich} P.,  {Tremaine} S.,  1978, ApJ, 222, 850

\bibitem[\protect\citeauthoryear{{Heinemann} \& {Papaloizou}}{{Heinemann} \&
  {Papaloizou}}{2009}]{HP09}
{Heinemann} T.,  {Papaloizou} J.~C.~B.,  2009, MNRAS, 397, 52

\bibitem[\protect\citeauthoryear{{Johnson} \& {Gammie}}{{Johnson} \&
  {Gammie}}{2003}]{JG03}
{Johnson} B.~M.,  {Gammie} C.~F.,  2003, ApJ, 597, 131

\bibitem[\protect\citeauthoryear{{Korycansky} \& {Pringle}}{{Korycansky} \&
  {Pringle}}{1995}]{KP95}
{Korycansky} D.~G.,  {Pringle} J.~E.,  1995, MNRAS, 272, 618

\bibitem[\protect\citeauthoryear{{Laughlin} \& {Bodenheimer}}{{Laughlin} \&
  {Bodenheimer}}{1994}]{LB94}
{Laughlin} G.,  {Bodenheimer} P.,  1994, ApJ, 436, 335

\bibitem[\protect\citeauthoryear{{Laughlin}, {Korchagin} \& {Adams}}{{Laughlin}
  et~al.}{1997}]{LKA97}
{Laughlin} G.,  {Korchagin} V.,    {Adams} F.~C.,  1997, ApJ, 477, 410

\bibitem[\protect\citeauthoryear{{Lin}, {Papaloizou} \& {Savonije}}{{Lin}
  et~al.}{1990}]{LPS90}
{Lin} D.~N.~C.,  {Papaloizou} J.~C.~B.,    {Savonije} G.~J.,  1990, ApJ, 364,
  326

\bibitem[\protect\citeauthoryear{{Lodato} \& {Rice}}{{Lodato} \&
  {Rice}}{2004}]{LR04}
{Lodato} G.,  {Rice} W.~K.~M.,  2004, MNRAS, 351, 630

\bibitem[\protect\citeauthoryear{{Lodato} \& {Rice}}{{Lodato} \&
  {Rice}}{2005}]{LR05}
{Lodato} G.,  {Rice} W.~K.~M.,  2005, MNRAS, 358, 1489

\bibitem[\protect\citeauthoryear{{Lubow} \& {Ogilvie}}{{Lubow} \&
  {Ogilvie}}{1998}]{LO98}
{Lubow} S.~H.,  {Ogilvie} G.~I.,  1998, ApJ, 504, 983

\bibitem[\protect\citeauthoryear{{Lubow} \& {Pringle}}{{Lubow} \&
  {Pringle}}{1993a}]{LP93b}
{Lubow} S.~H.,  {Pringle} J.~E.,  1993a, MNRAS, 263, 701

\bibitem[\protect\citeauthoryear{{Lubow} \& {Pringle}}{{Lubow} \&
  {Pringle}}{1993b}]{LP93a}
{Lubow} S.~H.,  {Pringle} J.~E.,  1993b, ApJ, 409, 360

\bibitem[\protect\citeauthoryear{{Mamatsashvili} \&
  {Chagelishvili}}{{Mamatsashvili} \& {Chagelishvili}}{2007}]{MC07}
{Mamatsashvili} G.~R.,  {Chagelishvili} G.~D.,  2007, MNRAS, 381, 809

\bibitem[\protect\citeauthoryear{{Mamatsashvili} \& {Rice}}{{Mamatsashvili} \&
  {Rice}}{2009}]{MR09}
{Mamatsashvili} G.~R.,  {Rice} W.~K.~M.,  2009, MNRAS, 394, 2153

\bibitem[\protect\citeauthoryear{{Mayer}, {Lufkin}, {Quinn} \&
  {Wadsley}}{{Mayer} et~al.}{2007}]{Mayeretal07}
{Mayer} L.,  {Lufkin} G.,  {Quinn} T.,    {Wadsley} J.,  2007, ApJ, 661, L77

\bibitem[\protect\citeauthoryear{{Mej{\'{\i}}a}, {Durisen}, {Pickett} \&
  {Cai}}{{Mej{\'{\i}}a} et~al.}{2005}]{Metal05}
{Mej{\'{\i}}a} A.~C.,  {Durisen} R.~H.,  {Pickett} M.~K.,    {Cai} K.,  2005,
  ApJ, 619, 1098

\bibitem[\protect\citeauthoryear{{Nelson}}{{Nelson}}{2006}]{N06}
{Nelson} A.~F.,  2006, MNRAS, 373, 1039

\bibitem[\protect\citeauthoryear{{Ogilvie}}{{Ogilvie}}{1998}]{Og98}
{Ogilvie} G.~I.,  1998, MNRAS, 297, 291

\bibitem[\protect\citeauthoryear{{Papaloizou} \& {Savonije}}{{Papaloizou} \&
  {Savonije}}{1991}]{PS91}
{Papaloizou} J.~C.,  {Savonije} G.~J.,  1991, MNRAS, 248, 353

\bibitem[\protect\citeauthoryear{{Papaloizou} \& {Lin}}{{Papaloizou} \&
  {Lin}}{1989}]{PL89}
{Papaloizou} J.~C.~B.,  {Lin} D.~N.~C.,  1989, ApJ, 344, 645

\bibitem[\protect\citeauthoryear{{Pickett}, {Cassen}, {Durisen} \&
  {Link}}{{Pickett} et~al.}{2000}]{Petal00}
{Pickett} B.~K.,  {Cassen} P.,  {Durisen} R.~H.,    {Link} R.,  2000, ApJ, 529,
  1034

\bibitem[\protect\citeauthoryear{{Pickett}, {Mej{\'{\i}}a}, {Durisen},
  {Cassen}, {Berry} \& {Link}}{{Pickett} et~al.}{2003}]{Petal03}
{Pickett} B.~K.,  {Mej{\'{\i}}a} A.~C.,  {Durisen} R.~H.,  {Cassen} P.~M.,
  {Berry} D.~K.,    {Link} R.~P.,  2003, ApJ, 590, 1060

\bibitem[\protect\citeauthoryear{{Rafikov}}{{Rafikov}}{2007}]{Raf07}
{Rafikov} R.~R.,  2007, ApJ, 662, 642

\bibitem[\protect\citeauthoryear{{Rice}, {Armitage}, {Bate} \&
  {Bonnell}}{{Rice} et~al.}{2003}]{Riceetal03}
{Rice} W.~K.~M.,  {Armitage} P.~J.,  {Bate} M.~R.,    {Bonnell} I.~A.,  2003,
  MNRAS, 339, 1025

\bibitem[\protect\citeauthoryear{{Rice}, {Lodato} \& {Armitage}}{{Rice}
  et~al.}{2005}]{RLA05}
{Rice} W.~K.~M.,  {Lodato} G.,    {Armitage} P.~J.,  2005, MNRAS, 364, L56

\bibitem[\protect\citeauthoryear{{Romeo}}{{Romeo}}{1992}]{Rom92}
{Romeo} A.~B.,  1992, MNRAS, 256, 307

\bibitem[\protect\citeauthoryear{{Romeo}}{{Romeo}}{1994}]{Rom94}
{Romeo} A.~B.,  1994, A\&A, 286, 799

\bibitem[\protect\citeauthoryear{{Ruden}, {Papaloizou} \& {Lin}}{{Ruden}
  et~al.}{1988}]{RPL88}
{Ruden} S.~P.,  {Papaloizou} J.~C.~B.,    {Lin} D.~N.~C.,  1988, ApJ, 329, 739

\bibitem[\protect\citeauthoryear{{Shu}}{{Shu}}{1968}]{Shu68}
{Shu} F.,  1968, PhD thesis, Harvard University, Cambridge, Massachusetts, USA

\bibitem[\protect\citeauthoryear{{Stamatellos} \& {Whitworth}}{{Stamatellos} \&
  {Whitworth}}{2009}]{SW09}
{Stamatellos} D.,  {Whitworth} A.~P.,  2009, MNRAS, 400, 1563

\bibitem[\protect\citeauthoryear{{Tevzadze}, {Chagelishvili} \&
  {Zahn}}{{Tevzadze} et~al.}{2008}]{TCZ08}
{Tevzadze} A.~G.,  {Chagelishvili} G.~D.,    {Zahn} J.,  2008, A\&A, 478, 9

\bibitem[\protect\citeauthoryear{{Toomre}}{{Toomre}}{1964}]{T64}
{Toomre} A.,  1964, ApJ, 139, 1217

\bibitem[\protect\citeauthoryear{{Toomre}}{{Toomre}}{1981}]{T81}
{Toomre} A.,  1981, in {S.~M.~Fall \& D.~Lynden-Bell} ed., Structure and
  Evolution of Normal Galaxies. Cambridge University Press, Cambridge

\end{thebibliography}

\end{document}